\documentclass[a4paper,UKenglish]{lipics-v2016}

\usepackage{amsmath}
\usepackage{xcolor}
\usepackage{listings}
\usepackage{graphicx}
\usepackage{booktabs}
\usepackage{mdwlist}
\usepackage{hyperref}
\usepackage{wrapfig}
\usepackage{xspace}

\usepackage[firstpage]{draftwatermark}
\SetWatermarkText{\hspace*{8in}\raisebox{8in}{\includegraphics[scale=0.1]{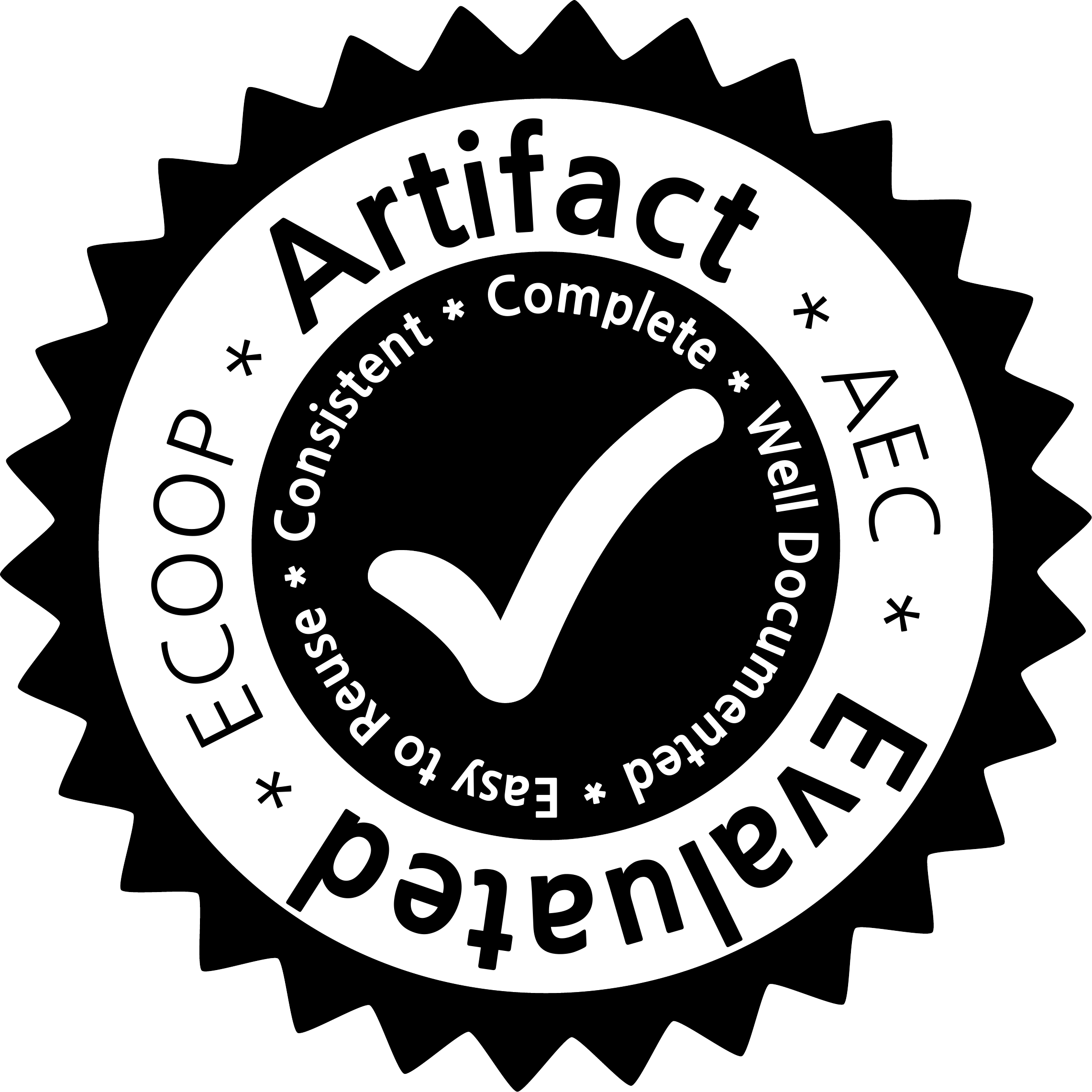}}}
\SetWatermarkAngle{0}

\definecolor{Blue}{RGB}{0,0,204}
\definecolor{Green}{RGB}{0,200,0}
\definecolor{Red}{RGB}{200,0,0}
\definecolor{Gray}{RGB}{150,150,150}

\newcommand{\hypothesis}[1]{Hypothesis #1}

\newcommand{\sqpytenoflags}[0]{$\mathrm{SQPyte}_{\mathrm{no-flags}}$\xspace}
\newcommand{\sqpytenoinline}[0]{$\mathrm{SQPyte}_{\mathrm{no-inline}}$\xspace}

\title{Making an Embedded DBMS JIT-friendly}

\author[1]{\href{http://cfbolz.de/}{Carl Friedrich Bolz}}
\author[2]{\href{http://cs.cmu.edu/\~dkurilov/}{Darya Kurilova}\thanks{Work performed on secondment at King's College London.}}
\author[1]{\href{http://tratt.net/laurie/}{Laurence Tratt}}
\affil[1]{Software Development Team, Department of Informatics, King's College London. \href{http://soft-dev.org/}{http://soft-dev.org/}}
\affil[2]{Institute for Software Research, School of Computer Science, Carnegie Mellon University.}

\authorrunning{C.\,F. Bolz, D. Kurilova, and L. Tratt}
\Copyright{Carl Friedrich Bolz, Darya Kurilova, and Laurence Tratt}

\subjclass{D.3.4 Processors}
\keywords{DBMSs, JIT, performance, tracing}

\EventEditors{??? and ???}
\EventNoEds{2}
\EventLongTitle{Proceedings of the 30th European Conference on Object-Oriented Programming (ECOOP 2016)}
\EventShortTitle{(ECOOP 2016)}
\EventAcronym{ECOOP}
\EventYear{2016}
\EventDate{July 2016}
\EventLocation{Rome, Italy}
\EventLogo{}
\SeriesVolume{1}
\ArticleNo{14}

\begin{document}

\maketitle

\begin{abstract}
While database management systems (DBMSs) are highly optimized,
interactions across the boundary between the programming language (PL) and the DBMS are costly,
even for in-process embedded DBMSs. In this paper, we show that programs
that interact with the popular embedded DBMS SQLite can be significantly
optimized -- by a factor of 3.4 in our benchmarks -- by inlining across
the PL / DBMS boundary. We achieved this speed-up by replacing parts of SQLite's
C interpreter with RPython code and composing the resulting meta-tracing virtual machine (VM) -- called
\emph{SQPyte} -- with the PyPy VM. SQPyte does not compromise stand-alone SQL
performance and is 2.2\% faster than SQLite on the widely used TPC-H benchmark
suite.
\end{abstract}

\section{Introduction}
\label{sec:introduction}

Significant effort goes into optimizing database management systems (DBMSs) and programming languages (PLs), and
both perform well in isolation: we can store and retrieve huge amounts of
complex data; and we can perform complex computations in reasonable time.
However, much less effort has gone into optimizing the interface between DBMSs
and programming languages. In some cases this is not surprising. Many DBMSs
run in separate processes -- and often on different computers -- to the PL calling them,
preventing meaningful optimisation across the two.
However, embedded DBMSs run in the same process as the PL
calling them and are thus potentially amenable to traditional PL optimisations.

In this paper, we aim to improve the performance of PLs that call
embedded DBMSs. Our fundamental hypothesis is the following:
\begin{quote}
  \textbf{Hypothesis 1} Optimisations that cross the barrier between a programming language
        and embedded DBMS significantly reduce the execution time of queries.
\end{quote}
In order to test this hypothesis, we composed together PyPy and
SQLite. PyPy is a widely used Python virtual machine (VM). SQLite is the most widely used
embedded DBMS, shipped by default with many operating systems, and used
by many applications. This composition required outfitting SQLite with a
Just-In-Time (JIT) compiler, which meant that we
also implicitly tested the following hypothesis:
\begin{quote}
  \textbf{Hypothesis 2} Replacing the query execution engine of a DBMS with
        a JIT reduces execution time of standalone SQL queries.
\end{quote}
Thus, we tested \hypothesis{2} before testing \hypothesis{1}. Our results strongly
validate \hypothesis{1} but, to our initial surprise, only weakly validate \hypothesis{2}.

The fundamental basis of the approach we took is to use
meta-tracing JIT compilers, as implemented by the RPython system. In essence,
from a description of an interpreter, RPython derives a VM
with a JIT compiler. PyPy is an existing RPython VM for the Python language.
SQLite, in contrast, is a traditional interpreter implemented in C. We 
therefore ported selected parts of SQLite's core opcode dispatcher from
C to RPython, turning SQLite into a (partially) meta-tracing DBMS. While we
left most of the core DBMS parts of SQLite (e.g.~B-tree manipulation, file
handling, and sorting) in C, we refer to our modified research system as \emph{SQPyte}
to simplify our exposition.

Relative to SQLite, SQPyte is $2.2\%$ faster on the industry standard TPC-H
benchmark suite~\cite{tpch}. We added specific optimisations intended to exploit the
fact that SQLite is dynamically typed, but, as this relatively paltry
performance improvement suggests, to little effect. We suspect that much more of
SQLite's C code would need to be ported to RPython for this figure to
significantly improve.

Since TPC-H measures SQL query performance in isolation from a
PL, we then created a series of micro-benchmarks which measure the performance
of programs which cross the PL / DBMS boundary. SQPyte is $3.4\times$ faster
than SQLite on these micro-benchmarks, showing the benefits of being able
to inline from PyPy into SQPyte.

The major parts of this paper are as follows. After describing how SQPyte was
created from SQLite (Section~\ref{sec:sqpyte}), we test Hypothesis 2
(Section~\ref{sec:evaluation1}). We then describe how PyPy and SQLite are
composed together (Section~\ref{sec:bridging}) allowing us to test Hypothesis 1
(Section~\ref{sec:evaluation2}).

SQPyte's source code, and all benchmarks used in this paper, can be downloaded from:

\begin{center}
\url{http://soft-dev.org/pubs/files/sqpyte/}
\end{center}

\section{Background}
\label{sec:background}

After briefly defining the difference between external and embedded databases,
this section summarizes the relevant aspects of SQLite and meta-tracing for those
readers who are unfamiliar with them. Note that this paper deals with several
different technologies, each of which uses slightly different
terminology. We have deliberately imposed consistent terminology in our
discussions to aid readers of this paper.

\subsection{Embedded DBMSs}

From the perspective of this paper, DBMSs come in two major variants.

\emph{External} DBMSs are typically used for large quantities of vital data.
They run as separate processes and interactions with them
require inter-process calls (IPCs) or network communications. The overhead of IPC varies depending on
operating system and hardware, but translating a function that returns a simple
integer into an IPC equivalent typically leads to a slowdown of at least 5
orders of magnitude. Since there is a fixed cost for each call, unrelated to the
quantity of data, small repeated IPC calls are costly. Programmers thus use
various techniques to bunch queries together to lower the fixed cost overhead of
IPC. When bunching is impossible, it is not unusual for IPC costs to dominate
interaction with an external DBMS. This effect is even more pronounced for
databases which run over a network.

\emph{Embedded} DBMSs are typically used for smaller
quantities of data, often as part of a desktop or mobile application.
They run within the same memory space
as a user program, removing IPC costs. However, embedded DBMSs tend to be
used as pre-packaged external libraries, meaning that there is no support
for optimising calls from the user application to the embedded DBMS.

\subsection{SQLite}

SQLite\footnote{\url{http://www.sqlite.org/}} is an embedded DBMS implemented
as a C library. It is the most commonly used embedded DBMS, installed as standard
on operating systems such as OS X, and widely used by desktop and mobile
applications (e.g.~email clients).

\begin{figure*}[t!]
\centering
\includegraphics[width=0.7\linewidth]{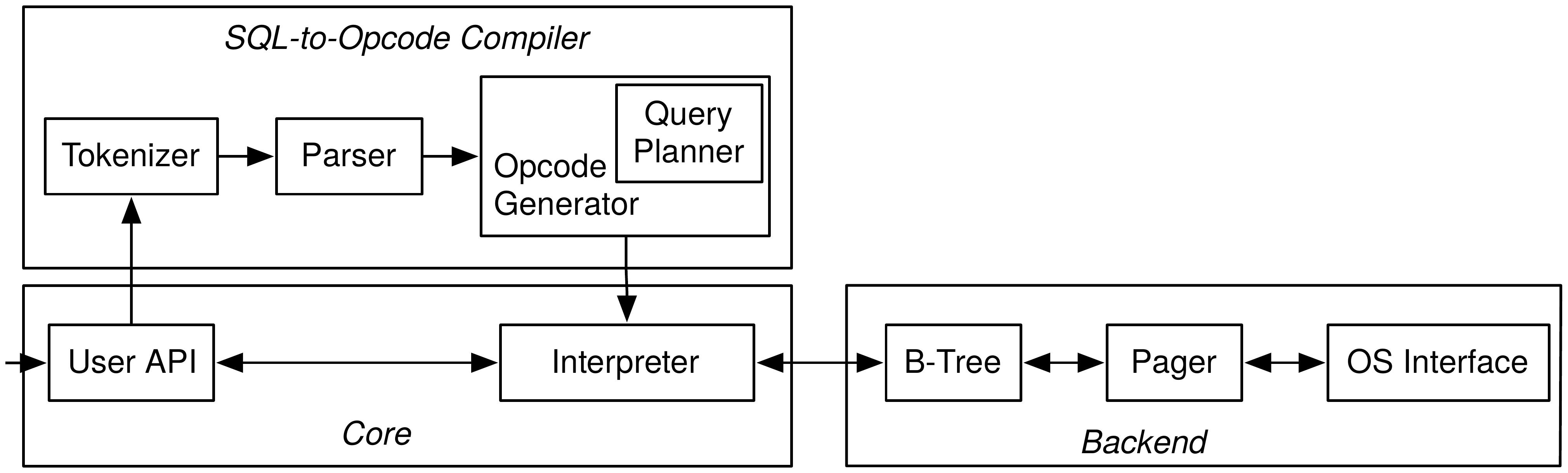}
\caption{SQLite architecture.}
\label{fig:sqlite-architecture}
\end{figure*}

Figure~\ref{fig:sqlite-architecture} shows SQLite's high-level
architecture: its core provides the user-facing API that external
programs use, as well as an interpreter for running queries; the backend stores and retrieves data
in memory and on disk; and the compiler translates SQL into an instruction
sequence. Instructions consist of an opcode (i.e.~the `type' of the instruction) and up to five operands
p1\ldots{}p5 (p1, p2, and p3 are always 32-bit integers; p4 is of variable size;
and p5 is an unsigned character), which are variously used to refer to registers,
program counter offsets, and the like. SQLite is dynamically typed and SQL values are
either Unicode strings, arbitrary binary `blobs', 64-bit numbers (integers or floating
point), or null. SQL values are stored in a single C-level type \texttt{Mem},
which can also store other SQLite internal values (e.g.~row sets). The opcode
dispatcher\footnote{SQLite refers to this as its `virtual machine', but we
reserve that term for other uses.} contains an arbitrary number of registers
(each of which stores a \texttt{Mem} instance), and zero or more cursors
(pointers into a table or index).

Figure~\ref{fig:mem-structure} shows an elided version of the \texttt{Mem} struct,
which plays a significant role in SQLite and therefore in much of our work.
\texttt{flags} is a bit field that
encodes the type(s) of the values stored in the struct. Most SQL values and
all SQLite internal values are stored in the \texttt{MemValue} union.
Strings are stored on the heap with a pointer to them in \texttt{z}.
In some cases, \texttt{Mem} can store two SQL values
simultaneously. For example, an integer cast at run-time to a string will store
the integer value in \texttt{i} and the string in \texttt{z}, so that
subsequent casts have zero cost. In such cases,
\texttt{flags} records that the value has more than one run-time type.

\begin{figure}
\begin{lstlisting}
struct Mem {
  union MemValue {
    double r;
    i64 i;
    ...
  } u;
  u16 flags;
  char *z;
  ...
};
\end{lstlisting}
\vspace{-10pt}
\caption{An elided view of SQLite's \lstinline{Mem} struct, used to represent
SQL values.}
\label{fig:mem-structure}
\end{figure}

\subsection{Meta-tracing}

Tracing is a technique for writing JIT compilers that record
`hot loops' in a program and convert them to machine code. Traditionally, tracing
requires manually creating both an interpreter and a trace compiler
(see~\cite{bala00dynamo,gal06hotpathvm}). In contrast, meta-tracing
takes an interpreter as input and from it automatically creates a
VM with a tracing JIT
compiler~\cite{mitchell70design,sullivan03dynamic,bolz09tracing,yermolovich09optimization,bebenita10spur}.
At run-time, user programs begin their execution in the
interpreter.  When a hot loop in the user program is encountered, the actions
of the interpreter are traced, optimized, and converted to
machine code. Since the initial traces are voluminous, the trace optimiser is
often able to reduce them to a fraction of their original size, before
they are converted to machine code.
Subsequent executions of the loop then use the fast machine code
version rather than the slow interpreter. Guards are left behind in the machine
code so that execution paths that stray from the trace revert back to the
interpreter.

In this paper we use RPython, the major extant meta-tracing language. RPython
is a statically typeable subset of Python with a type system similar to that of Java,
garbage collection, and high-level data types (e.g.~lists and dictionaries).
Despite this, VMs written in RPython have performance levels far exceeding traditional
interpreter-only implementations~\cite{bolz14impact}. The specific details
of RPython are generally unimportant in most of this paper, and we do not
concentrate on them: we believe that one could substitute any reasonable
meta-tracing language (or its cousin approach, self-optimizing interpreters with dynamic partial
evaluation~\cite{wuerthinger13onevm}) and achieve similar results.

\section{SQPyte}
\label{sec:sqpyte}

In order to test \hypothesis{2}, we created a variant of SQLite called
SQPyte, where parts of SQLite's interpreter are ported from C into RPython.
SQPyte is therefore meta-tracing compatible, meaning that SQL queries which use
the RPython parts of SQPyte's interpreter are JIT compiled. In the rest of this
section we explain the details of our porting.

It is important to note that SQPyte is not a complete, or even a majority,
rewrite of SQLite. Fortunately for us, RPython is compiled into C by default,
which makes mixing RPython and C code simple, with zero overhead for common
operations such as function calls. This means that we were able to leave most of
SQLite in C, calling such code as necessary, and only porting the minimum
possible to RPython. 

However, only code which is written in RPython can be JIT
compiled: calls to C cannot be inlined by the meta-tracer, reducing the
possibilities for
optimisations. We therefore worked incrementally, porting code to RPython
only after we had recognized that it was on the critical performance
path and likely to benefit from meta-tracing. Note that we are not suggesting
that SQPyte would not benefit from having more code in RPython, simply
that with our available effort levels, we had to focus our attention
on those parts of SQLite that we believed were most relevant.
As a rough indication of size,
we ported 1550 lines of C code and
wrote 1300 lines of RPython code to replace it.

In this section, we first introduce this paper's running example, before
explaining how SQPyte was created from SQLite.

\subsection{Running Example}
\label{sec:example}

\begin{figure}[t]
\begin{lstlisting}
import sqpyte
conn = sqpyte.Connection("tpch.db")
sum_qty = 0
sum_base_price = 0
sum_disc_price = 0
iterator = conn.execute("SELECT quantity, extendedprice, discount FROM lineitem")
for quantity, extendedprice, discount in iterator:
    sum_qty += quantity
    sum_base_price += extendedprice
    sum_disc_price += extendedprice * (1 - discount)
\end{lstlisting}
\caption{An example use of the \lstinline{sqpyte} module in PyPy. This
example program connects to the \lstinline{tpch.db} database and
computes the total quantity, the sum of the base price, and the sum of the
discounted price of all items.
balance of all accounts.}
\label{fig:sqpyte-example}
\end{figure}

SQPyte adds a module called \texttt{sqpyte} to PyPy. This module exposes a
standard Python DBMS API\footnote{The API is defined in
\url{https://www.python.org/dev/peps/pep-0249/}} that allows Python programs to
directly interact with SQPyte. Although it is mostly irrelevant from this
paper's perspective, the \texttt{sqpyte} module's interface is a strict
subset of that exposed by the \texttt{sqlite3} module, which is shipped
as standard with PyPy and other Python implementations.

Figure~\ref{fig:sqpyte-example} shows the running example we use
throughout this paper. After connecting to a database (line
2), the example starts the execution of an SQL query (line 6), receiving
an iterator object in return. As the iterator is pumped for new values
in the \texttt{for} loop (line 7), SQPyte lazily computes further values.
Each iteration yields 3 SQL values that are processed by regular
Python code (lines 8--10).

\subsection{Opcodes}

SQLite's interpreter executes instructions until either a query is complete, a
new row of results is produced, or
an error occurs. Each iteration of the interpreter loop loads the instruction at
the current program counter and jumps to an implementation of the instruction's
opcode.

\begin{figure*}[tb]
\vspace{-9px}
\hspace{-10pt}
\begin{tabular}{p{0.53\textwidth}p{0.47\textwidth}}
\begin{minipage}[t]{0.53\textwidth}
\begin{lstlisting}
SQLITE_PRIVATE int sqlite3VdbeExec(
   Vdbe *p) {
   int pc=0;                  
   Op *aOp = p->aOp;          
   Op *pOp;                   
   int rc = SQLITE_OK;        
   ...
   for(pc=p->pc; rc==SQLITE_OK; pc++){
     ...
     switch( pOp->opcode ){
       case OP_Goto: {             
         pc = pOp->p2 - 1;
         ...
         break;
       }
       case OP_Gosub: {            
         ...
         break;
       }
       case ...: { ... }
     }
   }
}
\end{lstlisting}
\end{minipage}
&
\hspace{-10ex}
\begin{minipage}[t]{0.47\textwidth}
\begin{lstlisting}
def mainloop(self):
  rc = CConfig.SQLITE_OK
  pc = self.p.pc
  while True:
    jitdriver.jit_merge_point(pc)
    if rc != CConfig.SQLITE_OK:
      break
    op = self._hlops[pc]
    opcode = op.get_opcode()
    oldpc = pc
    if opcode == CConfig.OP_Goto:
      pc, rc = self.python_OP_Goto(pc, rc, op)
    elif opcode == CConfig.OP_Gosub:
      pc = self.python_OP_Gosub(pc, op)
    elif ...:
      ...
    pc += 1
    if pc <= oldpc:
      jitdriver.can_enter_jit(pc)
\end{lstlisting}
\end{minipage}
\end{tabular}
\caption{An elided version of the opcode dispatcher, with the
original C on the left and the ported RPython on the right. The RPython interpreter
requires the \texttt{jit\_merge\_point} and \texttt{can\_enter\_jit} annotations
to enable the meta-tracing system to identify hot loops.}
\label{fig:interp_translation}
\end{figure*}

The first stage of the SQPyte port was to port the opcode dispatcher from C
to RPython, as shown in Figure~\ref{fig:interp_translation}. This can be thought
of as having three phases. First, since we wanted to reuse some of SQLite's
opcode's implementations, we split them out from the (rather large) switch statement they
were part of into individual functions (one per opcode). Second, we translated the main
opcode dispatcher loop itself. Finally, we added the two annotations\footnote{While these
are written using normal function call syntax, they are treated specially by RPython.}
required by RPython to make SQPyte's interpreter meta-tracing compatible. These
annotations inform the meta-tracing system about the current execution point of
the system (for example, the program counter and the known types of the
registers) so that it can determine if JIT compilation or
execution can, or should, occur. \texttt{can\_enter\_jit} is called when a loop
is encountered: if that happens often
enough, then tracing of the loop
occurs (i.e.~the loop is, ultimately, converted into machine code).
\texttt{jit\_merge\_point} allows the meta-tracing system to determine whether
there is a machine code version of the current execution point, or whether the
interpreter must be used instead.

\begin{figure*}[tb]
\vspace{-9px}
\begin{tabular}{p{0.53\textwidth}p{0.47\textwidth}}
\begin{minipage}[t]{\textwidth}
\begin{lstlisting}
case OP_IfPos: {
  pIn1 = &aMem[pOp->p1];
  assert(pIn1->flags&MEM_Int);
  VdbeBranchTaken(pIn1->u.i > 0, 2);
  if (pIn1->u.i > 0) {
     pc = pOp->p2 - 1;
  }
  break;
}
\end{lstlisting}
\end{minipage}
&
\hspace{-10ex}
\begin{minipage}[t]{\textwidth}
\begin{lstlisting}
def python_OP_IfPos(hlquery, pc, op):
    pIn1 = op.mem_of_p(1)
    if pIn1.get_u_i() > 0:
        pc = op.p2as_pc()
    return pc
\end{lstlisting}
\end{minipage}
\end{tabular}

\caption{An example port of an opcode from C to RPython. The \texttt{IfPos}
 opcode is a conditional jump: it loads the register specified by its p1 operand (lines 2 in C and RPython) and compares
the resulting value (as an integer) to 0 (line 5 in C; line 3 in RPython). If the value is greater
than zero it jumps to the position specified by the p2 operand (line 6
in C; line 4 in RPython). As this example shows, the RPython code makes greater
use of helper functions and removes functions that do not appear in the
production version of SQLite (both \texttt{assert} and \texttt{VdbeBranchTaken}
are no-ops in production builds).}
\label{fig:translation-example}
\end{figure*}

Figure~\ref{fig:translation-example} shows an example of an opcode in SQLite and
its SQPyte port. As this example suggests,
many aspects of the porting process are fairly obvious, though some are slightly
obscured by the greater use of helper functions in RPython (these make the
RPython version easier to understand in isolation, but can make C-to-RPython
comparisons a little harder). To ensure that we are able to make an
apples-to-apples performance comparison, we ported all aspects of SQLite's C code
enabled in the single-threaded default build. This meant that we did not need to port parts such as the
\texttt{assert} and \texttt{VdbeBranchTaken} macros (a complete list of unported 
aspects can be found in Appendix~\ref{sec:stuff we didn't port}), which are
no-ops in the default build and thus have no run-time effect whatsoever.

\begin{figure*}[t]
\vspace{-9px}
\begin{tabular}{p{0.53\textwidth}p{0.47\textwidth}}
\begin{minipage}[t]{\textwidth}
\begin{lstlisting}
case OP_MakeRecord: {
  ...
  if (...)
    goto no_mem;
  ...
}
case OP_Yield: {
  ...
}
...
no_mem:
  ...
\end{lstlisting}
\end{minipage}
&
\hspace{-10ex}
\begin{minipage}[t]{\textwidth}
\begin{lstlisting}
def OP_MakeRecord(...):
  ...
  if ...:
    return hlquery.gotoNoMem(pc)
  ...
def OP_Yield(...):
  ...
...
def gotoNoMem(hlquery, pc):
  ...
\end{lstlisting}
\end{minipage}
\end{tabular}
\caption{An example of how we port \texttt{goto}s in an opcode into SQPyte. We
ported 4 \texttt{goto} labels, making each a separate function
(e.g.~\texttt{gotoNoMem}). Instead of executing a \texttt{goto}, SQPyte calls
the appropriate function, and then immediately returns to the main interpreter
loop, thus mimicking the control flow of SQLite.}
\label{fig:dealing_with_gotos}
\end{figure*}

SQLite's opcode dispatcher contains several \texttt{goto}s to deal with
exceptional situations, as can be seen in
Figure~\ref{fig:dealing_with_gotos}. Since SQPyte breaks opcodes into different
functions, this behaviour is no longer tenable, since we can't \texttt{goto}
across different functions.\footnote{Not, it should be added, that RPython has a
\texttt{goto} construct.}
We thus ported labelled blocks to explicit functions, and \texttt{goto} jumps to
function calls, with each followed by a \texttt{return}. This achieves the
same overall program flow at the cost, when interpreting, of requiring more
function calls and, at any given time, an extra stack frame.

Porting all of SQLite's opcodes to RPython would be a significant task, and not
necessarily a fruitful one---some opcodes are called rarely, and some would
benefit little from meta-tracing. We thus chose to focus our porting efforts on
those opcodes which we believed would see the greatest benefit from meta-tracing
(chiefly those which change the program counter, or manipulate type flags).
Of SQLite's 153 opcodes, we ported 61 into RPython. A further 42 opcodes were
needed by queries we support, but we judged that they were unlikely to benefit from
JIT optimisations (because, for example, they immediately call SQLite's B-tree
manipulating
functions, which remain in C and are thus opaque to the meta-tracer). We thus
copied these opcodes directly from SQLite, leaving them in C. Since
we removed the giant \texttt{switch} statement these C opcodes were originally part of,
each was put into its own function, mirroring those opcodes ported to RPython.
Since this is a tedious, mechanical task, we copied only those opcodes
we needed: 50 of SQLite's opcodes are thus currently unsupported by
SQPyte, and an exception is raised if a query tries to use one of them.

\subsection{Optimizing the \texttt{flags} Attribute}
\label{sec:flags}

Most SQLite opcodes read or write to registers, each of which contains
a \texttt{Mem} struct. Typically, such opcodes
must first read the \texttt{flags} attribute of the \texttt{Mem} struct
to determine what type of value is stored within it. Many opcodes also
write to this flag when storing a result.
SQLite is completely dynamically typed -- different entries in a database
column, for example, may be of different types -- and, in essence,
the \texttt{flags} attribute is how the dynamic types are encoded. However, dynamically typed
languages tend to be surprisingly type-constant at run-time, which is why JIT
compilers are effective on such languages~\cite{bolz14impact}. A reasonable
expectation is thus that, as with other dynamically typed languages, most SQL
queries are fairly type-constant. We thus made the following hypothesis:
\begin{quote}
  \textbf{Hypothesis 3} Exposing the type information in the \texttt{flags} attribute
        associated with registers allows the JIT compiler to speed up query
        execution.
\end{quote}
We addressed this hypothesis by adding a mechanism to SQPyte that allows the trace optimiser
to reason about the \texttt{flags} attributes in registers' \texttt{Mem} structs.
This is implemented as a cache storing known \texttt{flags} values (in essence,
a close cousin of Self-style maps~\cite{chambers89efficient}).
When an opcode reads the \texttt{flags} attribute
from a \texttt{Mem} struct in a register, the trace records the read;
SQPyte is annotated such that the trace optimizer can remove all the subsequent
reads of the \texttt{flags} attribute of the same register, using the
previously read value. Similarly, subsequent reads are optimised away after a
\texttt{flags} attribute is written to. While the trace optimiser is normally
able to perform redundant load optimisations such as this automatically, it is
unable to reason about the \texttt{flags} attribute, which is stored in a
(semi-opaque) C object, hence our need to manually help the trace optimiser.

\begin{figure*}[t]
\vspace{-9px}
\hspace{-17pt}
\begin{tabular}{p{0.53\textwidth}p{0.47\textwidth}}
    \begin{minipage}[t]{\textwidth}
\begin{lstlisting}
case OP_NotNull: {
  pIn1 = &aMem[pOp->p1];
  VdbeBranchTaken((pIn1->flags & MEM_Null)==0, 2);
  if( (pIn1->flags & MEM_Null)==0 ){
    pc = pOp->p2 - 1;
  }
  break;
}
\end{lstlisting}
\end{minipage}
&
\hspace{13pt}
\begin{minipage}[t]{\textwidth}
\begin{lstlisting}
def OP_NotNull(hlquery, pc, op):
    pIn1, flags1 = op.mem_and_flags_of_p(1)
    if flags1 & CConfig.MEM_Null == 0:
        pc = op.p2as_pc()
    return pc
\end{lstlisting}
\end{minipage}
\end{tabular}

\caption{An example of porting operations on the \texttt{flags} attribute from C to
  RPython. In this case, the \texttt{NotNull} opcode jumps to a different
  \lstinline{pc} if the register indexed by the opcode's p1 operand is not
  \lstinline{Null}. The \lstinline{op.mem_and_flags_of_p(1)} helper function
  reads the register specified by the \texttt{p1} argument and returns the
  appropriate \texttt{Mem} structure and its flags.}
\label{fig:type-flags-example}
\end{figure*}

Figure~\ref{fig:type-flags-example} shows an example of the \texttt{NotNull}
opcode which operates on the \texttt{flags} attribute. The RPython method \lstinline{mem_and_flags_of_p()}
is the heart of the \texttt{flags} optimisation. If this opcode is part
of a trace which has earlier read \texttt{p1}'s flags, and there are no intermediate
writes, then the call to \lstinline{mem_and_flags_of_p(1)} will be entirely
removed by the trace optimizer.

\begin{figure}[t]
\begin{lstlisting}
@cache_safe(mutates="p2")
def python_OP_String(self, op):
  capi.impl_OP_String(...)
\end{lstlisting}
\caption{An example of the side-effect annotation used to specify which \texttt{flags}
attributes a C opcode can invalidate. In this case, the
annotation specifies that the \texttt{OP\_String} opcode invalidates the entry
for the register specified by the opcode's \texttt{p2} argument.}
\label{fig:effectinfo}
\end{figure}

Those opcodes which remain in C have their RPython wrapping function
annotated with side-effect information~\cite{Le:2005:UIS:2136624.2136653} to
specify which registers' flags may have been changed by the opcode.
After the opcode has been executed, the tracer knows that any previous information
about the \texttt{flags} fields of the relevant registers is now invalid. An
example annotation is shown in Figure~\ref{fig:effectinfo}.

Two opcodes are handled somewhat specially. First is SQLite's most frequently executed opcode,
\texttt{Column}, which reads one value from a row. This relatively complex opcode
analyses the packed B-Tree representing a row, extracts the requested column,
and stores it into the register specified by the \texttt{p3} operand. Because most of this opcode
calls out to DBMS C code, translating the entire (rather
large) opcode to RPython would be a tedious exercise.
Since the opcode can change register \texttt{p3}'s flags, this meant that most
calls to this opcode were followed by a check of \texttt{p3}'s
flags---including a read from memory. We removed these reads by
having the \texttt{Column} opcode return both the return code of the opcode and
the most recent value of \texttt{p3}'s \texttt{flags} encoded into one number.
The trace optimizer is then able to use the returned value to determine if its knowledge of \texttt{p3}'s
flags is current or not, without having to read from memory.

We also optimized the \texttt{MakeRecord} opcode to expose \texttt{flags} information to
the JIT compiler. This opcode reads from a specified number of $n$ registers and
produces a packed representation of the content of these registers, used for
later storage, and placed in the register specified by \texttt{p3}.
Since $n$ is constant for each
specific call of the opcode, we marked
\texttt{MakeRecord}'s inner loop as unrollable, so that the resulting
trace contains separate code for each register read. As well as removing
the general loop overhead, this allows the trace optimizer to reason about
the \texttt{flags} operations involved in reading from each of the $n$ registers.

\subsection{SQL Functions and Aggregates}

\begin{figure*}[htb]
\vspace{-9px}
\hspace{-11pt}
\begin{tabular}{p{0.53\textwidth}p{0.47\textwidth}}
    \begin{minipage}[t]{\textwidth}
\begin{lstlisting}
static void sin_sqlite(sqlite3_context *context,
       int argc, sqlite3_value **argv) {
    double value = sqlite3_value_double(argv[0]);
    double result = sin_sqlite(value);
    sqlite3_result_double(context, result);
}
...
sqlite3_create_function(db, "sin", 1, 
     SQLITE_UTF8, NULL, &sin_sqlite, NULL, NULL)
\end{lstlisting}
\end{minipage}
&
\hspace{7pt}
\begin{minipage}[t]{\textwidth}
\begin{lstlisting}
def sin(func, args, result):
    arg = args[0].sqlite3_value_double()
    result.sqlite3_result_double(math.sin(arg))
...
db.create_function("sin", 1, sin)
\end{lstlisting}
\end{minipage}
\end{tabular}
\caption{An example of registering a \texttt{sin} function with SQLite (C) and
SQPyte (RPython). Note that both APIs require specifying the name of
the function, the number of parameters, and a pointer to its implementation.}
\label{fig:function}
\end{figure*}

SQLite has both regular functions (henceforth simply
`functions') and aggregates.\footnote{There are also user-defined collation
functions which we did not optimize in a special way.} Both take a number of
arguments as input. Functions produce a single result per row that they are
applied to, whereas aggregates (e.g.~\lstinline{max}) reduce many rows to a
single value.

SQLite implements functions and aggregates in C, but does not hard-code
them into the interpreter: each is registered via an API to the SQL interpreter.
If SQPyte kept these functions in C, then the meta-tracer would have
to treat them as opaque calls, preventing inlining. Fortunately,
we were able to easily add an RPython mirror of SQLite's C interface
for registering functions and aggregates. Figure~\ref{fig:function} shows an
example of the two interfaces alongside each other. Aggregates are implemented
in similar manner, albeit in two parts: a step
function (e.g.~an acumulator) and a finalizer function (e.g.~a divisor).
We implemented a small number of commonly called SQL aggregates in
RPython: \lstinline{sum}, \lstinline{avg}, and \lstinline{count}.

To enable inlining, we also had to alter the opcodes which call functions
and aggregates. The \texttt{Function} opcode is responsible for calling
functions and is easily altered to permit inlining into RPython functions.
Calling an aggregate uses two opcodes: \texttt{AggStep} initializes the
aggregator, and calls the step function on each row;
and \texttt{AggFinalize} returns the final aggregate result.

\subsection{Overflow checking}

An advantage of controlling assembler code generation in a JIT is that one can
make use of machine code features that are hard to express directly in C.
RPython uses this to allow for overflow check's on arithmetic
operations to be performed without checking the operations concrete result
(i.e.~it makes use of hardware features which few programming languages
directly expose). We
make use of this feature in the implementation of arithmetic opcodes such as
\texttt{Add}, \texttt{Sub}, and \texttt{Mul}. If results overflow an integer,
each of these switch to a floating point representation.

\subsection{From Query to Trace}
\label{sec:running-sqlite}

We now recall the SQL query used in the running example of
Figure~\ref{fig:sqpyte-example}: \texttt{SELECT quantity, extendedprice,
discount FROM lineitem}. SQLite's compiler translates this into a sequence
of opcodes, which can be seen in Figure~\ref{fig:opcodes}.

\begin{figure*}[tb]
\vspace{-9px}
\hspace{-21pt}
\begin{tabular}{p{0.5\textwidth}p{0.5\textwidth}}
\begin{minipage}[t]{\textwidth}
{\begin{lstlisting}[numbers=none]
0|Init|0|12|0||00|
1|OpenRead|0|8|0|7|00|
2|Rewind|0|10|0||00|
3|Column|0|4|1||00|
4|Column|0|5|2||00|
5|RealAffinity|2|0|0||00|
6|Column|0|6|3||00|
7|RealAffinity|3|0|0||00|
8|ResultRow|1|3|0||00|
9|Next|0|3|0||01|
10|Close|0|0|0||00|
11|Halt|0|0|0||00|
12|Transaction|0|0|23|0|01|
13|TableLock|0|8|0|LineItem|00|
14|Goto|0|1|0||00|
\end{lstlisting}}
\end{minipage}
&
\hspace{-15pt}
\begin{minipage}[t]{\textwidth}
\begin{lstlisting}
# SQLite opcode Next
...
i168 = call(sqlite3BtreeNext, ...)
guard_value(i168, 0)
# SQLite opcode Column
i173 = call(impl_OP_Column, 3, ...)
guard_value(i173, 262144)
# SQLite opcode Column
i174 = call(impl_OP_Column, 4, ...)
guard_value(i174, 524288)
# SQLite opcode RealAffinity
# SQLite opcode Column
i175 = call(impl_OP_Column, 6, ...)
guard_value(i175, 524288)
# SQLite opcode RealAffinity
# SQLite opcode ResultRow
...
i178 = call(sqlite3VdbeCloseStatement, ...)
i179 = int_is_true(i178)
guard_false(i179)
...
\end{lstlisting}
\end{minipage}
\end{tabular}
\vspace{-9pt}
\caption{On the left, SQLite's rendering of the opcodes generated for the query \texttt{SELECT quantity,
extendedprice, discount FROM lineitem}. The first column represents the
program counter; the second column the opcode; and the remaining columns
the operands to the opcode. On the right, an elided SQPyte optimized trace for one result
row of the query. Note that after optimisation, some opcodes have no
operations in the trace.}
\vspace{-9pt}
\label{fig:opcodes}
\end{figure*}

The high-level structure of the query opcode as as follows. The query starts
by calling \texttt{Init} (opcode 0) which sets the program counter to its second operand, in this
case 12. This creates a new transaction (opcode 12), locks the table (opcode 13) before jumping (opcode 14) to the
main loop query.

The main loop operates on every row in the
database (opcodes 3--9). The \texttt{Column} opcodes (opcodes 3, 4, and 6) read
values from the \texttt{quantity}, \texttt{extendedprice}, and \texttt{discount}
columns in a row respectively. Although SQLite attaches type information to columns, these
are, in a sense, optional: any given value within a column may be of an arbitrary
type. Thus the \texttt{RealAffinity} opcodes (opcodes 5 and 7) inspect
the \texttt{extendedprice} and \texttt{discount} \texttt{Mem} structs: if they
hold floats (which SQLite terms `reals'), the result is a no-op; if they hold
integers, then they are cast to floats. The \texttt{ResultRow} opcode (opcode 8)
returns $n$ results (registers p1\ldots{}p1+p2-1 i.e.~1, 2, and 3 in our example) to the caller,
suspending query execution. Upon resumption, the \texttt{Next} opcode (opcode 9)
advances the database cursor to the next row in the table and updates
the program counter to its second operand -- in this case 3. If there is no
further data in the table, execution continues to the next opcode, which closes
the database connection (opcode 10) before halting query execution (opcode 11).

If the heart of the query opcode is in a hot loop traced by SQPyte's tracing JIT
compiler, then the result is as in Figure~\ref{fig:opcodes}. Traces always start
with the \texttt{Next} opcode, since the iteration that triggered the tracing
threshold was suspended as part of that opcode and thus the next iteration
starts when the query is resumed. \texttt{Next} calls the
\texttt{sqlite3BtreeNext} C function, which advances the database row (line 3),
with a guard ensuring the result is 0, which indicates success (line
4). The \texttt{Column} opcodes also call a C function, but the return type is
more complex, encoding both the function's error code and the flags of the
register that \texttt{Column} stored a result into. Assuming the guard holds,
the remainder of the trace thus implicitly knows the type of the register in
question (see Section~\ref{sec:flags}). This allows the trace optimizer to
remove the dynamic checks of the \texttt{RealAffinity} opcode all together. As
this shows, the trace optimizer is often able to remove a substantial portion of
the operations in an SQPyte trace.

\section{Experimental methodology}

We have two distinct experimental sections (primarily addressing, in order, Hypotheses 2
and 1), both sharing a common methodology. First we compare SQPyte to SQLite
and to H2, a widely used embedded Java database. H2 is of most interest to
Hypothesis 1, where it allows us to understand how SQPyte and PyPy's
cross-system inlining in RPython compares to Java and H2's cross-system inlining
on HotSpot. However, to put H2's cross-system performance into perspective, it
is also useful to see its performance on queries that address Hypothesis 2.
SQPyte is based on SQLite 3.8.8.2. We used PyPy 5.0 and
H2 1.4.191.

In both experimental sections, we run a number of queries. Each query is run in
5 fresh processes; each process runs 50 iterations of the query. We
placed a 1 hour timeout on each process. We report the
mean and 99\% confidence intervals of all iterations across all processes
(i.e.~250 in total). Note that by including all iterations, we are implicitly
including those where the VMs may be warming up.

As recommended by its documentation, SQLite was configured in single-threaded
mode, as was SQPyte. We used H2 in its default configuration. All benchmarks
were run on an otherwise idle Intel i7-4790 machine,
with 32 GiB RAM, and Debian 8.1. We turned off hyper-threading and turbo
boost in the BIOS: hyper-threading is of little use to our single-threaded
benchmarks, and adds noise to measurements; and turbo boost's benefits
disappear as soon as the CPU gets too hot, ruining benchmarking.
The database files were put into a RAM disk to ensure that possible
data caching effects between DBMSs were reduced. We performed an initial
run of our experiment to ensure that it never caused the machine to swap
memory to disk.

\section{Testing Hypothesis 2: SQPyte using TPC-H}
\label{sec:evaluation1}

\begin{figure}
\centering


\begin{tabular}{ l r@{\hskip 4pt} c@{\hskip 4pt} l r@{\hskip 4pt} c@{\hskip 4pt} l r@{\hskip 4pt} c@{\hskip 4pt} l }
\toprule
  \textbf{Benchmark} & \multicolumn{3}{ c }{\textbf{$\mathrm{SQPyte}$}} & \multicolumn{3}{ c }{\textbf{$\mathrm{SQLite}$}} & \multicolumn{3}{ c }{\textbf{$\mathrm{H2}$}} \\
\midrule
Query 1 (s) & 6.929 & \scriptsize{$\pm$} & \scriptsize{0.0352} & 8.715 & \scriptsize{$\pm$} & \scriptsize{0.0083} & 13.168 & \scriptsize{$\pm$} & \scriptsize{0.1584} \\
$\times$ &  &  &  & 1.258 & \scriptsize{$\pm$} & \scriptsize{0.0065} & 1.901 & \scriptsize{$\pm$} & \scriptsize{0.0254} \\
Query 2 (s) & 0.298 & \scriptsize{$\pm$} & \scriptsize{0.0098} & \textcolor{Gray}{0.305} & \textcolor{Gray}{\scriptsize{$\pm$}} & \textcolor{Gray}{\scriptsize{0.0024}} & 12.890 & \scriptsize{$\pm$} & \scriptsize{0.0787} \\
$\times$ &  &  &  & \textcolor{Gray}{1.025} & \textcolor{Gray}{\scriptsize{$\pm$}} & \textcolor{Gray}{\scriptsize{0.0340}} & 43.324 & \scriptsize{$\pm$} & \scriptsize{1.4273} \\
Query 3 (s) & 2.933 & \scriptsize{$\pm$} & \scriptsize{0.0329} & 3.098 & \scriptsize{$\pm$} & \scriptsize{0.0100} & 10.636 & \scriptsize{$\pm$} & \scriptsize{0.0490} \\
$\times$ &  &  &  & 1.056 & \scriptsize{$\pm$} & \scriptsize{0.0122} & 3.626 & \scriptsize{$\pm$} & \scriptsize{0.0452} \\
Query 4 (s) & 0.345 & \scriptsize{$\pm$} & \scriptsize{0.0038} & \textcolor{Gray}{0.345} & \textcolor{Gray}{\scriptsize{$\pm$}} & \textcolor{Gray}{\scriptsize{0.0014}} & 2.243 & \scriptsize{$\pm$} & \scriptsize{0.0265} \\
$\times$ &  &  &  & \textcolor{Gray}{0.998} & \textcolor{Gray}{\scriptsize{$\pm$}} & \textcolor{Gray}{\scriptsize{0.0121}} & 6.494 & \scriptsize{$\pm$} & \scriptsize{0.1081} \\
Query 5 (s) & 1.111 & \scriptsize{$\pm$} & \scriptsize{0.0145} & \textcolor{Gray}{1.116} & \textcolor{Gray}{\scriptsize{$\pm$}} & \textcolor{Gray}{\scriptsize{0.0239}} & 158.297 & \scriptsize{$\pm$} & \scriptsize{0.5371} \\
$\times$ &  &  &  & \textcolor{Gray}{1.004} & \textcolor{Gray}{\scriptsize{$\pm$}} & \textcolor{Gray}{\scriptsize{0.0261}} & 142.473 & \scriptsize{$\pm$} & \scriptsize{1.9971} \\
Query 6 (s) & 0.701 & \scriptsize{$\pm$} & \scriptsize{0.0081} & 0.794 & \scriptsize{$\pm$} & \scriptsize{0.0040} & 9.197 & \scriptsize{$\pm$} & \scriptsize{0.0571} \\
$\times$ &  &  &  & 1.134 & \scriptsize{$\pm$} & \scriptsize{0.0147} & 13.125 & \scriptsize{$\pm$} & \scriptsize{0.1741} \\
Query 7 (s) & 2.630 & \scriptsize{$\pm$} & \scriptsize{0.0070} & 2.847 & \scriptsize{$\pm$} & \scriptsize{0.0318} & 116.322 & \scriptsize{$\pm$} & \scriptsize{0.3302} \\
$\times$ &  &  &  & 1.083 & \scriptsize{$\pm$} & \scriptsize{0.0126} & 44.236 & \scriptsize{$\pm$} & \scriptsize{0.1710} \\
Query 8 (s) & 2.510 & \scriptsize{$\pm$} & \scriptsize{0.0141} & \textcolor{Gray}{2.519} & \textcolor{Gray}{\scriptsize{$\pm$}} & \textcolor{Gray}{\scriptsize{0.0646}} & 161.185 & \scriptsize{$\pm$} & \scriptsize{0.9576} \\
$\times$ &  &  &  & \textcolor{Gray}{1.003} & \textcolor{Gray}{\scriptsize{$\pm$}} & \textcolor{Gray}{\scriptsize{0.0265}} & 64.225 & \scriptsize{$\pm$} & \scriptsize{0.5471} \\
Query 9 (s) & 10.062 & \scriptsize{$\pm$} & \scriptsize{0.0448} & 10.269 & \scriptsize{$\pm$} & \scriptsize{0.0276} & 121.319 & \scriptsize{$\pm$} & \scriptsize{0.9515} \\
$\times$ &  &  &  & 1.021 & \scriptsize{$\pm$} & \scriptsize{0.0055} & 12.055 & \scriptsize{$\pm$} & \scriptsize{0.1137} \\
Query 10 (s) & 0.019 & \scriptsize{$\pm$} & \scriptsize{0.0056} & \textbf{0.009} & \textbf{\scriptsize{$\pm$}} & \textbf{\scriptsize{0.0006}} & 17.082 & \scriptsize{$\pm$} & \scriptsize{0.0660} \\
$\times$ &  &  &  & \textbf{0.499} & \textbf{\scriptsize{$\pm$}} & \textbf{\scriptsize{0.1632}} & 918.900 & \scriptsize{$\pm$} & \scriptsize{292.4240} \\
Query 11 (s) & 0.604 & \scriptsize{$\pm$} & \scriptsize{0.0071} & 0.647 & \scriptsize{$\pm$} & \scriptsize{0.0026} & \textbf{0.494} & \textbf{\scriptsize{$\pm$}} & \textbf{\scriptsize{0.0174}} \\
$\times$ &  &  &  & 1.071 & \scriptsize{$\pm$} & \scriptsize{0.0134} & \textbf{0.819} & \textbf{\scriptsize{$\pm$}} & \textbf{\scriptsize{0.0312}} \\
Query 12 (s) & 0.938 & \scriptsize{$\pm$} & \scriptsize{0.0062} & 1.027 & \scriptsize{$\pm$} & \scriptsize{0.0013} & 20.129 & \scriptsize{$\pm$} & \scriptsize{0.0604} \\
$\times$ &  &  &  & 1.094 & \scriptsize{$\pm$} & \scriptsize{0.0073} & 21.455 & \scriptsize{$\pm$} & \scriptsize{0.1571} \\
Query 13 (s) & 2.721 & \scriptsize{$\pm$} & \scriptsize{0.0135} & 2.818 & \scriptsize{$\pm$} & \scriptsize{0.0123} & 14.350 & \scriptsize{$\pm$} & \scriptsize{0.0840} \\
$\times$ &  &  &  & 1.036 & \scriptsize{$\pm$} & \scriptsize{0.0072} & 5.274 & \scriptsize{$\pm$} & \scriptsize{0.0427} \\
Query 14 (s) & 0.792 & \scriptsize{$\pm$} & \scriptsize{0.0102} & 0.863 & \scriptsize{$\pm$} & \scriptsize{0.0043} & 65.708 & \scriptsize{$\pm$} & \scriptsize{0.3407} \\
$\times$ &  &  &  & 1.090 & \scriptsize{$\pm$} & \scriptsize{0.0156} & 82.944 & \scriptsize{$\pm$} & \scriptsize{1.1772} \\
Query 15 (s) & 20.636 & \scriptsize{$\pm$} & \scriptsize{0.2254} & \textcolor{Gray}{20.881} & \textcolor{Gray}{\scriptsize{$\pm$}} & \textcolor{Gray}{\scriptsize{0.5542}} & \textbf{0.009} & \textbf{\scriptsize{$\pm$}} & \textbf{\scriptsize{0.0036}} \\
$\times$ &  &  &  & \textcolor{Gray}{1.011} & \textcolor{Gray}{\scriptsize{$\pm$}} & \textcolor{Gray}{\scriptsize{0.0300}} & \textbf{0.000} & \textbf{\scriptsize{$\pm$}} & \textbf{\scriptsize{0.0002}} \\
Query 16 (s) & 0.410 & \scriptsize{$\pm$} & \scriptsize{0.0074} & 0.447 & \scriptsize{$\pm$} & \scriptsize{0.0013} & 0.583 & \scriptsize{$\pm$} & \scriptsize{0.0155} \\
$\times$ &  &  &  & 1.089 & \scriptsize{$\pm$} & \scriptsize{0.0199} & 1.420 & \scriptsize{$\pm$} & \scriptsize{0.0459} \\
Query 17 (s) & 0.107 & \scriptsize{$\pm$} & \scriptsize{0.0008} & 0.114 & \scriptsize{$\pm$} & \scriptsize{0.0001} & 0.516 & \scriptsize{$\pm$} & \scriptsize{0.0141} \\
$\times$ &  &  &  & 1.067 & \scriptsize{$\pm$} & \scriptsize{0.0082} & 4.805 & \scriptsize{$\pm$} & \scriptsize{0.1354} \\
Query 18 (s) & 2.449 & \scriptsize{$\pm$} & \scriptsize{0.0144} & 2.822 & \scriptsize{$\pm$} & \scriptsize{0.0351} & 15.210 & \scriptsize{$\pm$} & \scriptsize{0.0745} \\
$\times$ &  &  &  & 1.152 & \scriptsize{$\pm$} & \scriptsize{0.0164} & 6.211 & \scriptsize{$\pm$} & \scriptsize{0.0492} \\
Query 19 (s) & 8.140 & \scriptsize{$\pm$} & \scriptsize{0.1333} & \textcolor{Gray}{8.114} & \textcolor{Gray}{\scriptsize{$\pm$}} & \textcolor{Gray}{\scriptsize{0.0397}} &  --- & &  \\
$\times$ &  &  &  & \textcolor{Gray}{0.997} & \textcolor{Gray}{\scriptsize{$\pm$}} & \textcolor{Gray}{\scriptsize{0.0169}} &  --- & &  \\
Query 20 (s) & 80.386 & \scriptsize{$\pm$} & \scriptsize{0.2692} & 81.378 & \scriptsize{$\pm$} & \scriptsize{0.2668} & \textbf{10.210} & \textbf{\scriptsize{$\pm$}} & \textbf{\scriptsize{0.0450}} \\
$\times$ &  &  &  & 1.012 & \scriptsize{$\pm$} & \scriptsize{0.0049} & \textbf{0.127} & \textbf{\scriptsize{$\pm$}} & \textbf{\scriptsize{0.0007}} \\
Query 21 (s) & 8.661 & \scriptsize{$\pm$} & \scriptsize{0.0347} & 9.066 & \scriptsize{$\pm$} & \scriptsize{0.1017} & \textbf{8.146} & \textbf{\scriptsize{$\pm$}} & \textbf{\scriptsize{0.0651}} \\
$\times$ &  &  &  & 1.047 & \scriptsize{$\pm$} & \scriptsize{0.0124} & \textbf{0.941} & \textbf{\scriptsize{$\pm$}} & \textbf{\scriptsize{0.0086}} \\
Query 22 (s) & 0.087 & \scriptsize{$\pm$} & \scriptsize{0.0036} & \textcolor{Gray}{0.087} & \textcolor{Gray}{\scriptsize{$\pm$}} & \textcolor{Gray}{\scriptsize{0.0003}} & 1.728 & \scriptsize{$\pm$} & \scriptsize{0.0215} \\
$\times$ &  &  &  & \textcolor{Gray}{1.003} & \textcolor{Gray}{\scriptsize{$\pm$}} & \textcolor{Gray}{\scriptsize{0.0408}} & 19.830 & \scriptsize{$\pm$} & \scriptsize{0.8475} \\
\midrule
\multicolumn{3}{ l }{\textbf{Geometric mean $\times$}} & & 1.022 & \scriptsize{$\pm$} & \scriptsize{0.0151} & 6.172 & \scriptsize{$\pm$} & \scriptsize{0.1484} \\
\bottomrule
\end{tabular}

\centering
\caption{SQPyte, SQLite, and H2 performance on the TPC-H benchmark
set. For each query, the first row shows the absolute time in seconds;
the second row shows the performance relative to SQPyte as a factor.
Queries where SQLite or H2 are faster than SQPyte are shown in bold.
Queries where SQLite or H2 are, within the confidence interval,
equivalent in performance to SQPyte are shown in grey. Note that Query
19 timed out on H2, hence the lack of data.}
\label{tab:tpch}
\end{figure}

To evaluate Hypothesis 2 -- in essence, does SQPyte have better performance than
SQLite when both are used standalone? -- we measure SQPyte's performance on the widely
used TPC-H benchmark set~\cite{tpch}. TPC-H's 22 queries utilise 8 tables,
which can be populated with different quantities of data: we chose the 1.5GiB
variant, which contains 8.7 million rows.
Figure~\ref{tab:tpch} shows the resulting comparison of the 3 DBMSs.

Overall, SQLite is $2.2\pm1.53\%$ slower than SQPyte. This validates Hypothesis
2, though only weakly. A more detailed look at the data reveals a slightly muddy
story. All but 1 query is faster in SQPyte than SQLite, with a maximum
improvement over SQLite of $25.8\pm0.65\%$ faster (query 1). Query 10 is the
outlier, with SQPyte a little over 100\%{} slower than SQLite. This is simply
because the query executes two orders of magnitude more quickly than all but one
other query ($0.0093\pm0.00061s$).
SQPyte's performance is thus dominated by
the time the JIT takes to produce machine code while in the first iteration of
the benchmark. However, even if query 10 were removed from the results, the
overall speedup would only be $5.8\pm0.45\%$ --- substantially better, but still
somewhat weak validation of Hypothesis 2. These results strongly suggest that for
benchmarks such as TPC-H, SQLite and SQPyte's overall performance is dominated
not by the interpreter but by the core DBMS (e.g.~operations on B-trees). Porting
more of SQLite to RPython may improve performance further, but it is hard to
estimate the likely gains, and the effort involved would be significant.

H2 is, on average, $6.172\pm0.1476\times$ significantly slower than both
SQPyte and SQLite. Query 19 exceeded our one hour timeout. Query 15, on the
other hand, is almost 3 orders of magnitude faster than SQLite and SQPyte. The
reason for that is that Query 15 uses an SQL view, which H2 is able to cache,
but which SQLite continually, and unnecessarily, recomputes (a well known SQLite issue).

\section{Composing SQPyte and PyPy}
\label{sec:bridging}

As with most embedded DBMSs, SQLite is rarely used standalone. Instead, a user
program interacts with SQLite through a language-specific library, as shown in
Figure~\ref{fig:sqpyte-example}. Thus the overall performance experienced by the
user is dictated by 3 factors: the performance of the programming language the
user program is implemented in; the performance of the embedded DBMS; and the
performance of interactions across the PL / DBMS boundary.
Hypothesis 1 captures our intuition that substantial optimisations are possible
if one can optimize across the PL / DBMS boundary.

In order to test Hypothesis 1, we composed together SQPyte and PyPy. PyPy is an
industrial strength meta-tracing Python VM, which can be used as a drop-in
replacement for the standard Python interpreter. Since PyPy is written in
RPython, we were able to extend SQPyte and PyPy so that tracing can
bridge across the PL / DBMS boundary. Put another way, database calls from
PyPy inline code in SQPyte's RPython interpreter.

The major part of the composition is the \texttt{sqpyte} module added to PyPy,
which allows programs run under PyPy to execute queries in SQPyte (see
Section~\ref{sec:example} for the user-facing details about this module). Since
it is written in RPython, \texttt{sqpyte} simply imports SQPyte as another
RPython module. Simple queries thus inline across the interface without
significant effort, with all the normal benefits of trace optimisation. The
optimisation of SQPyte's \texttt{flags} attribute (see Section~\ref{sec:flags})
means that in many cases data moved between SQPyte and PyPy requires
neither an explicit conversion nor even a guard. Some queries can't be sensibly
inlined, notably those which induce a loop in SQPyte's interpreter such as SQL
joins. In such
cases, PyPy and SQLite optimize their traces independently of each other.

\begin{figure*}[t]
\begin{lstlisting}
label(i144, f147, f154, i55, f57, f59, ...)
# for quantity, extendedprice, discount in iterator:
...
i161 = <MemValue 87403720>.u.i
f162 = <MemValue 87403776>.u.r
f163 = <MemValue 87403832>.u.r
...
# At this point, there is a copy of the trace from Figure 10
...
# sum_qty += quantity
i186 = int_add_ovf(i144, i161)
guard_no_overflow()

# sum_base_price += extendedprice
f188 = float_add(f147, f162)

# sum_disc_price += extendedprice * (1 - discount)
f189 = float_sub(1.000000, f163)
f190 = float_mul(f162, f189)
f191 = float_add(f154, f190)
...
jump(i186, f188, f191, i161, f162, f163, ...)
\end{lstlisting}
\caption{An elided version of the optimized trace of the Python program and SQL query from
Figure~\ref{fig:sqpyte-example}, annotated to explain which parts relate to
which parts of the input program. Notice that we have removed a significant part
of the trace at line 8, since it is identical to that found in
Figure~\ref{fig:opcodes}. As a rough gauge, the complete unoptimized trace
contains 375 operations; the optimized trace contains 137 operations.
}
\label{fig:pythontrace}
\end{figure*}

Using the running example from Figure~\ref{fig:sqpyte-example}, the resulting
trace in our composition can be seen in Figure~\ref{fig:pythontrace}. The
optimized trace
starts by reading the integer and two float values (\texttt{quantity},
\texttt{discount}, and \texttt{lineitem} respectively) from the \texttt{Mem}
structures of the most recently read row (lines 4--6). Next is a structurally
identical clone (with only $\alpha$-renamed SSA variables) of the trace from
Figure~\ref{fig:opcodes} (see the explanation in
Section~\ref{sec:running-sqlite}), which establishes the datatypes of the three
fetched values. The remainder of the trace (lines 10--22) correspond to the
Python \texttt{for} loop in Figure~\ref{fig:sqpyte-example}. Since the low-level
integer and float datatypes used by SQPyte and PyPy are the same, there is no
need to convert between the two, and, for example, the SQPyte integer (line 5)
can be used as-is in the PyPy part of the trace (line 11). Indeed, with the
exception of the overflow guard imposed by Python (line 12), the optimized trace
melds SQPyte and PyPy together such that it is difficult to distinguish the two.

\subsection{Calling back from SQPyte to Python}

\begin{figure}[t]
\begin{lstlisting}
class MySum(object):
    def __init__(self):
        self.sum = 0

    def step(self, x):
        self.sum += x

    def finalize(self):
        return self.sum
conn.create_aggregate("mysum", 1, MySum)
\end{lstlisting}
\caption{A pure Python implementation of a \texttt{sum} aggregate, registered
using \texttt{sqpyte}'s public API (line 10). Put another way, this example is
not part of SQPyte's RPython system, and is normal end-user code. For every
row of the query, the \texttt{step} method is called. The aggregation's
result is computed by calling the \texttt{finalize} method.}
\label{fig:pyaggregate}
\end{figure}

SQLite allows callbacks during an SQL query to functions in the calling PL.
For example, an end user can
register a new aggregate, which consumes a sequence of SQL rows and
returns a value as shown in Figure~\ref{fig:pyaggregate}. In our context,
Python can call SQLite, which calls Python, which returns to
SQLite, and which finally returns to Python. Since SQLite is reentrant, this
pattern of nesting can be arbitrarily deep.

While the ability to register such callbacks is powerful, it means that data and
control flow pass over the programming language / DBMS boundary much more
frequently than normal. The \texttt{sqpyte} module not only supports callback of
regular functions and aggregates, but enables inlining whenever possible.
Enabling this meant that we had to convert a few
more parts of SQLite into RPython, so that the full path from Python to SQPyte back to
Python is in RPython. Much as we did when calling SQPyte from Python, we
make use of tracings natural tendency to inline; though, as before, Python
callbacks which have loops lead to separate traces on either side.

\section{Evaluation of Hypothesis 1: SQPyte and PyPy}
\label{sec:evaluation2}

In this section we evaluate Hypothesis 1 -- in essence, does optimizing across
the boundary between PyPy and SQPyte lead to a significant performance increase?
-- and Hypothesis 3 -- in essence, does exposing type information in the
\texttt{flags} attribute increase performance? As well as benchmarking
SQLite, SQPyte, and H2 (as in Section~\ref{sec:evaluation1}), we also benchmark
two SQPyte variants: \sqpytenoinline turns off
inlining between SQPyte and PyPy and \sqpytenoflags turns off the type
flags optimisations of Section~\ref{sec:flags}.

\subsection{Micro-benchmarks for PyPy Integration}

\begin{figure}[t]
\begin{lstlisting}
d = {}
for key, suppkey in conn.execute("""SELECT PartSupp.PartKey, PartSupp.SuppKey
                                    FROM PartSupp;"""):
    cursor = conn.execute("SELECT Part.name FROM Part WHERE part.PartKey = ?;", [key])
    partname, = cursor.next()
    cursor = conn.execute("""SELECT Supplier.name FROM Supplier
                             WHERE Supplier.SuppKey = ?;""", [suppkey])
    suppname, = cursor.next()
    d[partname] = suppname
return d
\end{lstlisting}
\caption{The core of the \texttt{pythonjoin} micro-benchmark.}
\label{fig:pythonjoin}
\end{figure}

The TPC-H benchmarks measure SQL performance in isolation, but tell us
nothing about the performance of a PL calling a DBMS. Indeed,
to the best of our knowledge, there are no relevant benchmarks in this style.
In order to test Hypothesis 1, we were therefore forced to create 6 micro-benchmarks,
each designed to pass large quantities of data
across the PL / DBMS boundary. While they are not necessarily completely
realistic programs, they exemplify common idioms in larger programs
(see Figures~\ref{fig:sqpyte-example} and \ref{fig:pythonjoin}). The
micro-benchmarks are as follows:
\begin{description}
  \item[select] is the running example of Figure~\ref{fig:sqpyte-example}.
      The DBMS query iterates over three columns of a table, returning them
      to Python, which performs arithmetic operations on the results.
    \item[innerjoin] joins 3 tables with an inner join and returns the
        resulting tuples to Python, which are then stored into a hashmap.
    \item[pythonjoin] implements a semantically equivalent join to the
       \emph{innerjoin} benchmark, but does so in Python rather than using the DBMS. The
        Python code iterates over 1 of the tables, on each iteration executing
        2 sub-queries for the other 2 tables.
        On the Python side the tuples are stored into a hashmap. The
        core part of this micro-benchmark can be seen in Figure~\ref{fig:pythonjoin}.
    \item[pyfunction] models calling back to a Python function from SQL.
        An \texttt{abs} function is defined in pure Python. The SQL query
        then iterates over all rows in a table, calling \texttt{abs} on
        one column, and returning that column's value to Python, which then
        sums all the elements.
    \item[pyaggregate] models calling an aggregate defined in Python. A
        \texttt{sum} aggregate is defined in pure Python (as in Figure~\ref{fig:pyaggregate})
        and used to sum one column of a table.
    \item[filltable] first adds 100,000 rows to a two-column table, with
      each row being added in a single SQL query. A single SQL query
      then reads all of the added rows back out again.
\end{description}
Each benchmark has a Java equivalent such that we can run it with H2.
All micro-benchmarks use the TPC-H dataset from Section~\ref{sec:evaluation1},
with the exception of the \texttt{filltable} micro-benchmark which creates,
writes, and reads from its own tables. Figure~\ref{tab:crossings} shows how
often each micro-benchmark crosses between DBMS and PL, and how many values are
converted between the DBMS and PL.

\begin{figure}
\centering
\begin{tabular}{lrr}
\toprule
\textbf{Benchmark} & \textbf{Crossings} & \textbf{Values converted} \\
\midrule
select      &  6\,001\,217 & 18\,003\,645 \\
innerjoin   &     800\,002 &  1\,600\,000 \\
pythonjoin  &  4\,000\,002 &  4\,800\,000 \\
pyfunction  & 12\,002\,434 & 18\,003\,645 \\
pyaggregate &  6\,001\,218 & 12\,002\,431 \\
filltable   &     200\,004 &     400\,000 \\
\bottomrule
\end{tabular}
\centering
\caption{How often the micro-benchmarks cross the boundary between Python and
the database, and how many values are converted across the boundary in
total. In most cases, the `values converted' is a whole-number multiple of
`crossings'. \texttt{pyfunction} crosses the PL / DBMS boundary twice
per iteration, with one crossing returning one value, the other two.
\texttt{pythonjoin} has a similar, though more complex, pattern of crossings
to \texttt{pyfunction}.}
\label{tab:crossings}
\end{figure}

\subsection{Results and Evaluation}
\label{sec:microbenchmarks evaluation}

\begin{figure}
\centering


\begin{tabular}{ l r@{\hskip 4pt} c@{\hskip 4pt} l r@{\hskip 4pt} c@{\hskip 4pt} l r@{\hskip 4pt} c@{\hskip 4pt} l }
\toprule
  \textbf{Benchmark} & \multicolumn{3}{ c }{\textbf{$\mathrm{SQPyte}$}} & \multicolumn{3}{ c }{\textbf{$\mathrm{SQLite}$}} & \multicolumn{3}{ c }{\textbf{$\mathrm{H2}$}} \\
\midrule
select (s) & 0.772 & \scriptsize{$\pm$} & \scriptsize{0.0081} & 3.382 & \scriptsize{$\pm$} & \scriptsize{0.0114} & 73.095 & \scriptsize{$\pm$} & \scriptsize{0.9489} \\
$\times$ &  &  &  & 4.382 & \scriptsize{$\pm$} & \scriptsize{0.0515} & 94.662 & \scriptsize{$\pm$} & \scriptsize{1.6645} \\
innerjoin (s) & 0.578 & \scriptsize{$\pm$} & \scriptsize{0.0030} & 0.913 & \scriptsize{$\pm$} & \scriptsize{0.0032} & 20.957 & \scriptsize{$\pm$} & \scriptsize{0.1636} \\
$\times$ &  &  &  & 1.579 & \scriptsize{$\pm$} & \scriptsize{0.0102} & 36.268 & \scriptsize{$\pm$} & \scriptsize{0.3472} \\
pythonjoin (s) & 1.397 & \scriptsize{$\pm$} & \scriptsize{0.0061} & 3.332 & \scriptsize{$\pm$} & \scriptsize{0.0862} & 9.292 & \scriptsize{$\pm$} & \scriptsize{0.0776} \\
$\times$ &  &  &  & 2.385 & \scriptsize{$\pm$} & \scriptsize{0.0585} & 6.651 & \scriptsize{$\pm$} & \scriptsize{0.0628} \\
pyfunction (s) & 0.580 & \scriptsize{$\pm$} & \scriptsize{0.0027} & 3.861 & \scriptsize{$\pm$} & \scriptsize{0.0930} & 55.298 & \scriptsize{$\pm$} & \scriptsize{0.3916} \\
$\times$ &  &  &  & 6.661 & \scriptsize{$\pm$} & \scriptsize{0.1678} & 95.402 & \scriptsize{$\pm$} & \scriptsize{0.8443} \\
pyaggregate (s) & 0.542 & \scriptsize{$\pm$} & \scriptsize{0.0289} & 2.558 & \scriptsize{$\pm$} & \scriptsize{0.0712} & 8.218 & \scriptsize{$\pm$} & \scriptsize{0.0387} \\
$\times$ &  &  &  & 4.730 & \scriptsize{$\pm$} & \scriptsize{0.2893} & 15.167 & \scriptsize{$\pm$} & \scriptsize{0.7735} \\
filltable (s) & 0.067 & \scriptsize{$\pm$} & \scriptsize{0.0013} & 0.188 & \scriptsize{$\pm$} & \scriptsize{0.0149} & 1.565 & \scriptsize{$\pm$} & \scriptsize{0.0443} \\
$\times$ &  &  &  & 2.805 & \scriptsize{$\pm$} & \scriptsize{0.2336} & 23.342 & \scriptsize{$\pm$} & \scriptsize{0.8382} \\
\midrule
\multicolumn{3}{ l }{\textbf{Geometric mean $\times$}} & & 3.367 & \scriptsize{$\pm$} & \scriptsize{0.0648} & 30.283 & \scriptsize{$\pm$} & \scriptsize{0.3563} \\
\bottomrule
\end{tabular}

\centering
\caption{Results of the micro-benchmark set. The table shows absolute times
in seconds, as well as the relative factor of each VM normalized to SQPyte.
The last row contains the geometric mean of the normalized factors.
Micro-benchmarks where SQLite or H2 are faster than SQPyte are shown in bold.
Micro-benchmarks where SQLite or H2 are, within the confidence interval,
equivalent in performance to SQPyte are shown in grey.}
\label{tab:micro}
\end{figure}

The results of the micro-benchmarks are shown in Figure~\ref{tab:micro}. They
show that on these conversion-heavy queries SQPyte outperforms SQLite by a
factor of $3.367\pm0.0637\times$ on average. Figure~\ref{tab:crossings} shows
how often each micro-benchmark crosses the DBMS / PL boundary. As predicted by
Hypothesis 1, the more often a micro-benchmark crosses the boundary (as shown in
Figure~\ref{tab:crossings}), the greater SQPyte's advantage.

H2, in contrast, is significantly slower on
these benchmarks -- $30.285\pm0.3515\times$  -- than on the TPC-H benchmarks.
We believe that this is because HotSpot is unable to optimize effectively across
the PL / DBMS boundary. Unfortunately, definitively verifying that this is the
cause is impossible, as we cannot selectively turn on and off the relevant
HotSpot optimisations. However, the magnitude of the effect strongly
suggests that simply having both PL and DBMS running on the same VM is not
sufficient to optimize across the PL / DBMS boundary effectively.

In order to understand the cause of SQPyte's good performance on the
micro-benchmarks, we created \sqpytenoinline, a simple variant of SQPyte which
disables all inlining between SQPyte and PyPy. Note that although no inlining
occurs, traces in \sqpytenoinline are still created on both sides of the PL /
DBMS boundary, so we are able to make a sensible comparison between SQPyte and
\sqpytenoinline.

The resulting figures are shown in the second columns of
Figures~\ref{tab:micro-analysis} and~\ref{tab:tpch-analysis}. As expected, there
is no statistical difference in the performance of SQPyte and \sqpytenoinline on
the TPC-H benchmarks ($0.3\pm1.99\%$)---the only Python code in these benchmarks
is that used to consume the results of a query, ensuring that the database
definitely produces the results.

The micro-benchmarks are rather different, with \sqpytenoinline being
$2.388\pm0.0350\times$ slower than SQPyte. This shows that inlining is the
single biggest part of the speed benefit of SQPyte relative to SQLite. The only
micro-benchmark that is relatively little affected is \texttt{innerjoin}, which
is $1.266\pm0.0079\times$ slower than SQPyte. This is because most of the work
in the benchmark is involved in the table joins, which happen entirely in the
DBMS. In contrast, \texttt{pyfunction}, which crosses the boundary twice per iteration (once
from Python to the DBMS, and then from the query calling back to Python) sees a
large slowdown in \sqpytenoinline of $3.501\pm0.0555\times$.

In summary, not only does SQPyte give a significant performance increase when
the DBMS / PL boundary is crossed regularly, but we can see that inlining is the
major factor in this. This strongly validates \hypothesis{1}.

\section{Evaluation of Hypothesis 3: The Effect of Optimizing the \texttt{flags} Attribute}

\begin{figure}
\centering


\begin{tabular}{ l r@{\hskip 4pt} c@{\hskip 4pt} l r@{\hskip 4pt} c@{\hskip 4pt} l r@{\hskip 4pt} c@{\hskip 4pt} l }
\toprule
  \textbf{Benchmark} & \multicolumn{3}{ c }{\textbf{$\mathrm{SQPyte}$}} & \multicolumn{3}{ c }{\textbf{$\mathrm{SQPyte}_{\mathrm{no-inline}}$}} & \multicolumn{3}{ c }{\textbf{$\mathrm{SQPyte}_{\mathrm{no-flags}}$}} \\
\midrule
select (s) & 0.772 & \scriptsize{$\pm$} & \scriptsize{0.0081} & 1.948 & \scriptsize{$\pm$} & \scriptsize{0.0190} & 0.795 & \scriptsize{$\pm$} & \scriptsize{0.0029} \\
$\times$ &  &  &  & 2.524 & \scriptsize{$\pm$} & \scriptsize{0.0383} & 1.030 & \scriptsize{$\pm$} & \scriptsize{0.0119} \\
innerjoin (s) & 0.578 & \scriptsize{$\pm$} & \scriptsize{0.0030} & 0.732 & \scriptsize{$\pm$} & \scriptsize{0.0025} & \textcolor{Gray}{0.579} & \textcolor{Gray}{\scriptsize{$\pm$}} & \textcolor{Gray}{\scriptsize{0.0017}} \\
$\times$ &  &  &  & 1.266 & \scriptsize{$\pm$} & \scriptsize{0.0079} & \textcolor{Gray}{1.003} & \textcolor{Gray}{\scriptsize{$\pm$}} & \textcolor{Gray}{\scriptsize{0.0060}} \\
pythonjoin (s) & 1.397 & \scriptsize{$\pm$} & \scriptsize{0.0061} & 2.961 & \scriptsize{$\pm$} & \scriptsize{0.0796} & 1.423 & \scriptsize{$\pm$} & \scriptsize{0.0117} \\
$\times$ &  &  &  & 2.117 & \scriptsize{$\pm$} & \scriptsize{0.0605} & 1.019 & \scriptsize{$\pm$} & \scriptsize{0.0099} \\
pyfunction (s) & 0.580 & \scriptsize{$\pm$} & \scriptsize{0.0027} & 2.029 & \scriptsize{$\pm$} & \scriptsize{0.0302} & 0.605 & \scriptsize{$\pm$} & \scriptsize{0.0021} \\
$\times$ &  &  &  & 3.501 & \scriptsize{$\pm$} & \scriptsize{0.0551} & 1.044 & \scriptsize{$\pm$} & \scriptsize{0.0062} \\
pyaggregate (s) & 0.542 & \scriptsize{$\pm$} & \scriptsize{0.0289} & 1.380 & \scriptsize{$\pm$} & \scriptsize{0.0619} & \textcolor{Gray}{0.529} & \textcolor{Gray}{\scriptsize{$\pm$}} & \textcolor{Gray}{\scriptsize{0.0023}} \\
$\times$ &  &  &  & 2.547 & \scriptsize{$\pm$} & \scriptsize{0.1862} & \textcolor{Gray}{0.976} & \textcolor{Gray}{\scriptsize{$\pm$}} & \textcolor{Gray}{\scriptsize{0.0498}} \\
filltable (s) & 0.067 & \scriptsize{$\pm$} & \scriptsize{0.0013} & 0.206 & \scriptsize{$\pm$} & \scriptsize{0.0017} & \textcolor{Gray}{0.069} & \textcolor{Gray}{\scriptsize{$\pm$}} & \textcolor{Gray}{\scriptsize{0.0016}} \\
$\times$ &  &  &  & 3.072 & \scriptsize{$\pm$} & \scriptsize{0.0670} & \textcolor{Gray}{1.032} & \textcolor{Gray}{\scriptsize{$\pm$}} & \textcolor{Gray}{\scriptsize{0.0327}} \\
\midrule
\multicolumn{3}{ l }{\textbf{Geometric mean $\times$}} & & 2.388 & \scriptsize{$\pm$} & \scriptsize{0.0346} & 1.017 & \scriptsize{$\pm$} & \scriptsize{0.0113} \\
\bottomrule
\end{tabular}

\centering
\caption{Results of the micro-benchmark set. \sqpytenoinline disables inlining
across the database-programming language boundary, \sqpytenoflags disables the
optimisation that reasons about the \texttt{flags} attribute of the \texttt{Mem}
structures. The table shows average absolute times
in seconds, as well as the factor of SQPyte normalized to each of the other
VMs. The last row contains the geometric mean of the normalized factors. 
Micro-benchmarks where SQLite or H2 are faster than SQPyte are shown in bold.
Micro-benchmarks where SQLite or H2 are, within the confidence interval,
equivalent in performance to SQPyte are shown in grey.}
\label{tab:micro-analysis}
\end{figure}

\begin{figure}
\centering


\begin{tabular}{ l r@{\hskip 4pt} c@{\hskip 4pt} l r@{\hskip 4pt} c@{\hskip 4pt} l r@{\hskip 4pt} c@{\hskip 4pt} l }
\toprule
  \textbf{Benchmark} & \multicolumn{3}{ c }{\textbf{$\mathrm{SQPyte}$}} & \multicolumn{3}{ c }{\textbf{$\mathrm{SQPyte}_{\mathrm{no-inline}}$}} & \multicolumn{3}{ c }{\textbf{$\mathrm{SQPyte}_{\mathrm{no-flags}}$}} \\
\midrule
Query 1 (s) & 6.929 & \scriptsize{$\pm$} & \scriptsize{0.0352} & \textcolor{Gray}{6.903} & \textcolor{Gray}{\scriptsize{$\pm$}} & \textcolor{Gray}{\scriptsize{0.0137}} & 7.082 & \scriptsize{$\pm$} & \scriptsize{0.0239} \\
$\times$ &  &  &  & \textcolor{Gray}{0.996} & \textcolor{Gray}{\scriptsize{$\pm$}} & \textcolor{Gray}{\scriptsize{0.0055}} & 1.022 & \scriptsize{$\pm$} & \scriptsize{0.0064} \\
Query 2 (s) & 0.298 & \scriptsize{$\pm$} & \scriptsize{0.0098} & \textcolor{Gray}{0.296} & \textcolor{Gray}{\scriptsize{$\pm$}} & \textcolor{Gray}{\scriptsize{0.0065}} & \textbf{0.283} & \textbf{\scriptsize{$\pm$}} & \textbf{\scriptsize{0.0039}} \\
$\times$ &  &  &  & \textcolor{Gray}{0.995} & \textcolor{Gray}{\scriptsize{$\pm$}} & \textcolor{Gray}{\scriptsize{0.0385}} & \textbf{0.953} & \textbf{\scriptsize{$\pm$}} & \textbf{\scriptsize{0.0345}} \\
Query 3 (s) & 2.933 & \scriptsize{$\pm$} & \scriptsize{0.0329} & \textcolor{Gray}{2.922} & \textcolor{Gray}{\scriptsize{$\pm$}} & \textcolor{Gray}{\scriptsize{0.0136}} & \textcolor{Gray}{2.957} & \textcolor{Gray}{\scriptsize{$\pm$}} & \textcolor{Gray}{\scriptsize{0.0379}} \\
$\times$ &  &  &  & \textcolor{Gray}{0.996} & \textcolor{Gray}{\scriptsize{$\pm$}} & \textcolor{Gray}{\scriptsize{0.0124}} & \textcolor{Gray}{1.008} & \textcolor{Gray}{\scriptsize{$\pm$}} & \textcolor{Gray}{\scriptsize{0.0184}} \\
Query 4 (s) & 0.345 & \scriptsize{$\pm$} & \scriptsize{0.0038} & \textcolor{Gray}{0.346} & \textcolor{Gray}{\scriptsize{$\pm$}} & \textcolor{Gray}{\scriptsize{0.0038}} & \textcolor{Gray}{0.346} & \textcolor{Gray}{\scriptsize{$\pm$}} & \textcolor{Gray}{\scriptsize{0.0041}} \\
$\times$ &  &  &  & \textcolor{Gray}{1.002} & \textcolor{Gray}{\scriptsize{$\pm$}} & \textcolor{Gray}{\scriptsize{0.0159}} & \textcolor{Gray}{1.001} & \textcolor{Gray}{\scriptsize{$\pm$}} & \textcolor{Gray}{\scriptsize{0.0169}} \\
Query 5 (s) & 1.111 & \scriptsize{$\pm$} & \scriptsize{0.0145} & \textcolor{Gray}{1.114} & \textcolor{Gray}{\scriptsize{$\pm$}} & \textcolor{Gray}{\scriptsize{0.0117}} & \textcolor{Gray}{1.097} & \textcolor{Gray}{\scriptsize{$\pm$}} & \textcolor{Gray}{\scriptsize{0.0176}} \\
$\times$ &  &  &  & \textcolor{Gray}{1.003} & \textcolor{Gray}{\scriptsize{$\pm$}} & \textcolor{Gray}{\scriptsize{0.0185}} & \textcolor{Gray}{0.987} & \textcolor{Gray}{\scriptsize{$\pm$}} & \textcolor{Gray}{\scriptsize{0.0221}} \\
Query 6 (s) & 0.701 & \scriptsize{$\pm$} & \scriptsize{0.0081} & \textcolor{Gray}{0.693} & \textcolor{Gray}{\scriptsize{$\pm$}} & \textcolor{Gray}{\scriptsize{0.0016}} & \textcolor{Gray}{0.702} & \textcolor{Gray}{\scriptsize{$\pm$}} & \textcolor{Gray}{\scriptsize{0.0014}} \\
$\times$ &  &  &  & \textcolor{Gray}{0.990} & \textcolor{Gray}{\scriptsize{$\pm$}} & \textcolor{Gray}{\scriptsize{0.0118}} & \textcolor{Gray}{1.001} & \textcolor{Gray}{\scriptsize{$\pm$}} & \textcolor{Gray}{\scriptsize{0.0117}} \\
Query 7 (s) & 2.630 & \scriptsize{$\pm$} & \scriptsize{0.0070} & \textcolor{Gray}{2.637} & \textcolor{Gray}{\scriptsize{$\pm$}} & \textcolor{Gray}{\scriptsize{0.0178}} & 2.657 & \scriptsize{$\pm$} & \scriptsize{0.0105} \\
$\times$ &  &  &  & \textcolor{Gray}{1.003} & \textcolor{Gray}{\scriptsize{$\pm$}} & \textcolor{Gray}{\scriptsize{0.0075}} & 1.010 & \scriptsize{$\pm$} & \scriptsize{0.0049} \\
Query 8 (s) & 2.510 & \scriptsize{$\pm$} & \scriptsize{0.0141} & \textcolor{Gray}{2.494} & \textcolor{Gray}{\scriptsize{$\pm$}} & \textcolor{Gray}{\scriptsize{0.0167}} & \textcolor{Gray}{2.499} & \textcolor{Gray}{\scriptsize{$\pm$}} & \textcolor{Gray}{\scriptsize{0.0270}} \\
$\times$ &  &  &  & \textcolor{Gray}{0.994} & \textcolor{Gray}{\scriptsize{$\pm$}} & \textcolor{Gray}{\scriptsize{0.0088}} & \textcolor{Gray}{0.996} & \textcolor{Gray}{\scriptsize{$\pm$}} & \textcolor{Gray}{\scriptsize{0.0124}} \\
Query 9 (s) & 10.062 & \scriptsize{$\pm$} & \scriptsize{0.0448} & \textcolor{Gray}{10.102} & \textcolor{Gray}{\scriptsize{$\pm$}} & \textcolor{Gray}{\scriptsize{0.0227}} & \textcolor{Gray}{10.138} & \textcolor{Gray}{\scriptsize{$\pm$}} & \textcolor{Gray}{\scriptsize{0.0766}} \\
$\times$ &  &  &  & \textcolor{Gray}{1.004} & \textcolor{Gray}{\scriptsize{$\pm$}} & \textcolor{Gray}{\scriptsize{0.0051}} & \textcolor{Gray}{1.007} & \textcolor{Gray}{\scriptsize{$\pm$}} & \textcolor{Gray}{\scriptsize{0.0090}} \\
Query 10 (s) & 0.019 & \scriptsize{$\pm$} & \scriptsize{0.0056} & \textcolor{Gray}{0.020} & \textcolor{Gray}{\scriptsize{$\pm$}} & \textcolor{Gray}{\scriptsize{0.0057}} & \textcolor{Gray}{0.023} & \textcolor{Gray}{\scriptsize{$\pm$}} & \textcolor{Gray}{\scriptsize{0.0089}} \\
$\times$ &  &  &  & \textcolor{Gray}{1.079} & \textcolor{Gray}{\scriptsize{$\pm$}} & \textcolor{Gray}{\scriptsize{0.4796}} & \textcolor{Gray}{1.237} & \textcolor{Gray}{\scriptsize{$\pm$}} & \textcolor{Gray}{\scriptsize{0.6278}} \\
Query 11 (s) & 0.604 & \scriptsize{$\pm$} & \scriptsize{0.0071} & \textcolor{Gray}{0.602} & \textcolor{Gray}{\scriptsize{$\pm$}} & \textcolor{Gray}{\scriptsize{0.0093}} & \textcolor{Gray}{0.600} & \textcolor{Gray}{\scriptsize{$\pm$}} & \textcolor{Gray}{\scriptsize{0.0073}} \\
$\times$ &  &  &  & \textcolor{Gray}{0.998} & \textcolor{Gray}{\scriptsize{$\pm$}} & \textcolor{Gray}{\scriptsize{0.0205}} & \textcolor{Gray}{0.994} & \textcolor{Gray}{\scriptsize{$\pm$}} & \textcolor{Gray}{\scriptsize{0.0178}} \\
Query 12 (s) & 0.938 & \scriptsize{$\pm$} & \scriptsize{0.0062} & \textcolor{Gray}{0.937} & \textcolor{Gray}{\scriptsize{$\pm$}} & \textcolor{Gray}{\scriptsize{0.0067}} & \textcolor{Gray}{0.934} & \textcolor{Gray}{\scriptsize{$\pm$}} & \textcolor{Gray}{\scriptsize{0.0045}} \\
$\times$ &  &  &  & \textcolor{Gray}{0.999} & \textcolor{Gray}{\scriptsize{$\pm$}} & \textcolor{Gray}{\scriptsize{0.0099}} & \textcolor{Gray}{0.995} & \textcolor{Gray}{\scriptsize{$\pm$}} & \textcolor{Gray}{\scriptsize{0.0083}} \\
Query 13 (s) & 2.721 & \scriptsize{$\pm$} & \scriptsize{0.0135} & 2.767 & \scriptsize{$\pm$} & \scriptsize{0.0420} & \textcolor{Gray}{2.767} & \textcolor{Gray}{\scriptsize{$\pm$}} & \textcolor{Gray}{\scriptsize{0.0461}} \\
$\times$ &  &  &  & 1.017 & \scriptsize{$\pm$} & \scriptsize{0.0164} & \textcolor{Gray}{1.017} & \textcolor{Gray}{\scriptsize{$\pm$}} & \textcolor{Gray}{\scriptsize{0.0178}} \\
Query 14 (s) & 0.792 & \scriptsize{$\pm$} & \scriptsize{0.0102} & \textcolor{Gray}{0.785} & \textcolor{Gray}{\scriptsize{$\pm$}} & \textcolor{Gray}{\scriptsize{0.0048}} & \textcolor{Gray}{0.793} & \textcolor{Gray}{\scriptsize{$\pm$}} & \textcolor{Gray}{\scriptsize{0.0101}} \\
$\times$ &  &  &  & \textcolor{Gray}{0.991} & \textcolor{Gray}{\scriptsize{$\pm$}} & \textcolor{Gray}{\scriptsize{0.0151}} & \textcolor{Gray}{1.001} & \textcolor{Gray}{\scriptsize{$\pm$}} & \textcolor{Gray}{\scriptsize{0.0196}} \\
Query 15 (s) & 20.636 & \scriptsize{$\pm$} & \scriptsize{0.2254} & \textcolor{Gray}{20.437} & \textcolor{Gray}{\scriptsize{$\pm$}} & \textcolor{Gray}{\scriptsize{0.1008}} & \textcolor{Gray}{20.674} & \textcolor{Gray}{\scriptsize{$\pm$}} & \textcolor{Gray}{\scriptsize{0.3404}} \\
$\times$ &  &  &  & \textcolor{Gray}{0.991} & \textcolor{Gray}{\scriptsize{$\pm$}} & \textcolor{Gray}{\scriptsize{0.0120}} & \textcolor{Gray}{1.002} & \textcolor{Gray}{\scriptsize{$\pm$}} & \textcolor{Gray}{\scriptsize{0.0210}} \\
Query 16 (s) & 0.410 & \scriptsize{$\pm$} & \scriptsize{0.0074} & \textcolor{Gray}{0.416} & \textcolor{Gray}{\scriptsize{$\pm$}} & \textcolor{Gray}{\scriptsize{0.0088}} & \textcolor{Gray}{0.443} & \textcolor{Gray}{\scriptsize{$\pm$}} & \textcolor{Gray}{\scriptsize{0.0380}} \\
$\times$ &  &  &  & \textcolor{Gray}{1.014} & \textcolor{Gray}{\scriptsize{$\pm$}} & \textcolor{Gray}{\scriptsize{0.0295}} & \textcolor{Gray}{1.079} & \textcolor{Gray}{\scriptsize{$\pm$}} & \textcolor{Gray}{\scriptsize{0.0972}} \\
Query 17 (s) & 0.107 & \scriptsize{$\pm$} & \scriptsize{0.0008} & \textcolor{Gray}{0.107} & \textcolor{Gray}{\scriptsize{$\pm$}} & \textcolor{Gray}{\scriptsize{0.0007}} & \textcolor{Gray}{0.108} & \textcolor{Gray}{\scriptsize{$\pm$}} & \textcolor{Gray}{\scriptsize{0.0008}} \\
$\times$ &  &  &  & \textcolor{Gray}{1.001} & \textcolor{Gray}{\scriptsize{$\pm$}} & \textcolor{Gray}{\scriptsize{0.0102}} & \textcolor{Gray}{1.009} & \textcolor{Gray}{\scriptsize{$\pm$}} & \textcolor{Gray}{\scriptsize{0.0112}} \\
Query 18 (s) & 2.449 & \scriptsize{$\pm$} & \scriptsize{0.0144} & \textcolor{Gray}{2.441} & \textcolor{Gray}{\scriptsize{$\pm$}} & \textcolor{Gray}{\scriptsize{0.0043}} & 2.485 & \scriptsize{$\pm$} & \scriptsize{0.0145} \\
$\times$ &  &  &  & \textcolor{Gray}{0.997} & \textcolor{Gray}{\scriptsize{$\pm$}} & \textcolor{Gray}{\scriptsize{0.0062}} & 1.015 & \scriptsize{$\pm$} & \scriptsize{0.0091} \\
Query 19 (s) & 8.140 & \scriptsize{$\pm$} & \scriptsize{0.1333} & \textcolor{Gray}{8.082} & \textcolor{Gray}{\scriptsize{$\pm$}} & \textcolor{Gray}{\scriptsize{0.0744}} & \textbf{7.883} & \textbf{\scriptsize{$\pm$}} & \textbf{\scriptsize{0.0611}} \\
$\times$ &  &  &  & \textcolor{Gray}{0.993} & \textcolor{Gray}{\scriptsize{$\pm$}} & \textcolor{Gray}{\scriptsize{0.0191}} & \textbf{0.968} & \textbf{\scriptsize{$\pm$}} & \textbf{\scriptsize{0.0182}} \\
Query 20 (s) & 80.386 & \scriptsize{$\pm$} & \scriptsize{0.2692} & \textcolor{Gray}{80.801} & \textcolor{Gray}{\scriptsize{$\pm$}} & \textcolor{Gray}{\scriptsize{0.3028}} & \textcolor{Gray}{80.542} & \textcolor{Gray}{\scriptsize{$\pm$}} & \textcolor{Gray}{\scriptsize{0.3804}} \\
$\times$ &  &  &  & \textcolor{Gray}{1.005} & \textcolor{Gray}{\scriptsize{$\pm$}} & \textcolor{Gray}{\scriptsize{0.0054}} & \textcolor{Gray}{1.002} & \textcolor{Gray}{\scriptsize{$\pm$}} & \textcolor{Gray}{\scriptsize{0.0061}} \\
Query 21 (s) & 8.661 & \scriptsize{$\pm$} & \scriptsize{0.0347} & \textcolor{Gray}{8.684} & \textcolor{Gray}{\scriptsize{$\pm$}} & \textcolor{Gray}{\scriptsize{0.0782}} & \textbf{8.572} & \textbf{\scriptsize{$\pm$}} & \textbf{\scriptsize{0.0340}} \\
$\times$ &  &  &  & \textcolor{Gray}{1.003} & \textcolor{Gray}{\scriptsize{$\pm$}} & \textcolor{Gray}{\scriptsize{0.0101}} & \textbf{0.990} & \textbf{\scriptsize{$\pm$}} & \textbf{\scriptsize{0.0059}} \\
Query 22 (s) & 0.087 & \scriptsize{$\pm$} & \scriptsize{0.0036} & \textcolor{Gray}{0.087} & \textcolor{Gray}{\scriptsize{$\pm$}} & \textcolor{Gray}{\scriptsize{0.0037}} & \textcolor{Gray}{0.084} & \textcolor{Gray}{\scriptsize{$\pm$}} & \textcolor{Gray}{\scriptsize{0.0020}} \\
$\times$ &  &  &  & \textcolor{Gray}{0.998} & \textcolor{Gray}{\scriptsize{$\pm$}} & \textcolor{Gray}{\scriptsize{0.0601}} & \textcolor{Gray}{0.959} & \textcolor{Gray}{\scriptsize{$\pm$}} & \textcolor{Gray}{\scriptsize{0.0465}} \\
\midrule
\multicolumn{3}{ l }{\textbf{Geometric mean $\times$}} & & \textcolor{Gray}{1.003} & \textcolor{Gray}{\scriptsize{$\pm$}} & \textcolor{Gray}{\scriptsize{0.0197}} & \textcolor{Gray}{1.010} & \textcolor{Gray}{\scriptsize{$\pm$}} & \textcolor{Gray}{\scriptsize{0.0237}} \\
\bottomrule
\end{tabular}

\centering
\caption{Further variants of SQPyte running TPC-H. For a description of the
columns see Figure~\ref{tab:micro-analysis}.}
\label{tab:tpch-analysis}
\end{figure}

In order to see how much the optimisation of the \texttt{flags} attribute of the
\texttt{Mem} struct described in Section~\ref{sec:flags} helps, we created a
version \sqpytenoflags of SQPyte that disables this optimisation completely and
reran all benchmarks. The results are shown in the last columns of
Figures~\ref{tab:micro-analysis} and~\ref{tab:tpch-analysis}. 

We expected that turning off the \texttt{flags} optimisations would slow
execution down, and that it would account for much of the performance benefit
not accounted for by inlining in Section~\ref{sec:microbenchmarks evaluation}.
On the TPC-H benchmarks, there is no statistically
observable effect (a slowdown $1.0\pm2.41\%$). On the micro-benchmarks,
the slowdown is statistically significant ($1.7\pm1.112\%$), but only
very marginally.

These results were not what we expected, and lead us to reject
Hypothesis 3.

\section{Threats to Validity}
\label{sec:threats to validity}

Benchmarks can only provide a partial view of a system's overall performance,
and thus don't necessarily reflect the behaviour of more realistic settings and
workloads. The TPC-H benchmarks are widely used,
though the micro-benchmarks are our own creations, and we may unintentionally
have created micro-benchmarks which unduly flatter SQPyte.

When porting SQLite's interpreter to SQPyte, we only ported those parts enabled
in the default build of SQLite. Since some parts are tangled up in C
\texttt{\#{}ifdef}s, we may have unintentionally misclassified one or more of
these parts. Appendix~\ref{sec:stuff we didn't port} contains a complete list of
the parts we did not port, so that readers can verify our choices.

There is a subtle difference between PyPy calling SQLite and SQPyte: in the former
case, PyPy uses a C FFI (the \texttt{cffi} module in PyPy) to interface with
SQLite; in the latter, PyPy simply imports the SQPyte system as an RPython module. There is
thus the potential of additional
overhead when PyPy calls SQLite compared to when it calls SQPyte. We examined
the PyPy traces for the case when it calls SQLite, and verified that the
overhead is extremely small (a small handful of machine
code instructions), and insignificant relative to the difference to the two
systems.

\section{Related Work}
\label{sec:related}

We split our discussion of related work into two sections: optimizing SQL
with code generation; and optimizing the interactions between PLs and DBMSs.

\subsection{Optimizing Execution SQL with Code Generation}

Many databases use the \emph{iterator model} for query
execution~\cite{Lorie1974-jp, graefe_volcano_1993} which, in essence,
is equivalent to an AST interpreter in PL implementation.
There have been many attempts to generate
code from query plans to reduce the overheads of the iterator model. This
started with very early databases such as System R~\cite{chamberlin_history_1981}. Most of these
approaches require code generators to be written by hand. In contrast,
SQPyte's meta-tracing JIT compiler implicitly
implements the semantics of the system by tracing the RPython interpreter.

Rao et al.~\cite{rao06} describe a relational, Java-based, in-memory database that, for each query, dynamically 
generates new query-specific code. They created two versions of the query
planner: an interpreted one using the iterator model and a compiled one. They
demonstrated that using the compiled version removed 
the overhead of virtual functions in the interpreted version. In addition, the
Java JIT compiler was much better at optimizing the generated code for each
query than the interpreted version. On average, the compiled queries in their
benchmark ran twice as fast as the interpreter.

Krikellas et al.~\cite{krikellas10} generate C code from queries and load the
compiled shared libraries to execute them.
Their compilation process dynamically instantiates carefully handwritten
templates to create source code specific for a given query and hardware.
The performance of their dynamically generated evaluation plans are comparable to hard-coded equivalents.

Neumann~\cite{neumann11} describes an approach where the query is compiled to
machine code using the LLVM compiler~\cite{lattner_2004}. When generating code, the
approach attempts to keep data in registers for as long as possible.
Similar to how SQPyte interacts with the existing SQLite C code, this system
also preserves complex parts of the database in C++ and calls into the C++ code
from the LLVM code as needed. The resulting system is $2-4\times$
faster than the other databases benchmarked.

Klonatos et al.~\cite{klonatos14,Klonatos:2014:EBE:2733004.2733084,Rompf:2015:FPS:2784731.2784760}~use generative programming techniques in
Scala to dynamically compile queries using a Scala SQL query engine into C.
This technique implicitly inlines parts of the DBMS into the query
and shows good performance on the TPC-H benchmark suite relative to the
DBX DBMS. However, because code written in Scala cannot inline the generated C,
there is no equivalent of the PyPy / SQPyte bridge we implemented.

Haensch et al.~\cite{hansch_2015} describe an automatic operator model
specialization system. Their system uses a general LLVM-based specialization component to
dynamically optimize the execution of query plans in the operator model with
the help of partial evaluation. Certain query operator fields are marked
as immutable, allowing the specializer to
aggressively inline and optimize the query plan execution. This approach
suffers somewhat from having to perform its optimisations at the (rather
low-level) LLVM IR, at which point a lot of useful high-level information has
been lost. Both this approach and SQPyte use interpreters as the basis of the
run-time code generation, though Haensch et al.'s interpreters are closer in
style to the AST-based partial evaluation system of
Truffle~\cite{wuerthinger13onevm}.

\subsection{Optimizing Language-Database Interaction}

Grust et al.~\cite{grust09} created the Ferry glue language which serves as an
intermediate language to translate subsets of other languages to SQL. That is
various front-end languages can translate to that intermediate language,
which is then lowered into SQL code. The goal is to reduce the impedance
mismatch between languages and database when programming and to improve the
efficiency of the interaction.
Ferry influenced several works in other languages: Garcia et al.~\cite{garcia10}
developed a Scala plugin that enables programmers to translate Scala-level
constructs to Ferry;
Grust et al.~\cite{grust12} introduced Switch which uses Ferry-like translation principles to allow seamless integration 
of Ruby and Ruby on Rails with the DBMS; Giorgidze et al.~\cite{giorgidze10} designed and implemented a Haskell library 
for database-supported program execution; and Schreiber et al.~\cite{schreiber10} created a Ferry-based
LINQ-to-SQL provider.
Since one of the main goals of Ferry is to reduce the number of times the
PL / DBMS boundary is crossed, the approach is complementary to the
SQPyte approach of reducing the cost of the boundary crossings.

Mattis et al.~\cite{mattis_columnar_2015} describe columnar objects, which is an
implementation of an in-memory column store embedded into Python together with a
seamless integration into the Python object system. With the help
of the PyPy JIT compiler they produce efficient machine code for Python code
that queries the data store. Compared to SQPyte their approach offers a much
deeper integration of the database implementation into the host language, at the
cost of having to implement the data store from scratch.

Unipycation by Barrett et
al.~\cite{barrett_bolz_tratt__approaches_to_interpreter_composition} is a
language composition of Prolog and Python that uses meta-tracing to reduce the
overhead of crossing the boundary between the two languages. It composes
together PyPy with Pyrolog, a Prolog interpreter written in
RPython. As with SQPyte, the most effective optimisation is inlining.

\section{Conclusion}
\label{sec:conclusion}

This paper's major result is that there are substantial, and previously
missed, opportunities for optimizing across the PL / DBMS boundary.
We achieved a significant performance increase by inlining queries from
SQL into PyPy. Furthermore, most of
this performance increase came from tracing's natural tendency to inline---our
attempts to add more complex dynamic typing optimisations had little effect.

Those who wish to apply our approach to other embedded DBMSs can take heart from
this: a relatively simple conversion of parts of an interpreter implemented in C into a
meta-tracing language is highly effective.
We estimate that we spent at least 8 person-months on the SQPyte implementation, with
perhaps half of that spent on the \texttt{flags} optimisation, and a
further 2 person-months on the PyPy / SQPyte bridge. For this
relatively moderate effort -- certainly compared to the much greater work put
into both SQLite and PyPy -- we were able to substantially improve performance
for queries that regularly cross the PL / DBMS boundary. While we were only able
to marginally increase the performance of stand-alone SQL queries, we did not
encounter any examples where SQPyte is slower than SQLite. This suggests that
SQPyte, or a similar system based on SQLite, may be useful to a wider range
of users.

Our approach of incrementally replacing SQLite's C code with RPython had an
interesting trade-off. It made initial development easier, since we always had a running
system. However, it had disadvantages which became more apparent in later
stages of development. Most obviously, since it necessitated keeping core data-structures in C, we
hobbled the trace optimizer somewhat. The best -- or, from our perspective, worst --
example of this is the \texttt{flags} optimisation which, despite significant
effort, ended up slightly slowing our system down. We suspect that
porting more of these data-structures, and the code that relies on them, into RPython would enable further
performance increases. Indeed, were we to tackle SQPyte from scratch, we
might place less emphasis on keeping interpreter data-structures in C---we conjecture
that in several places we might have incurred less effort on our part if
we had ported more C data-structures into RPython.

A secondary, and largely implicit result, is that we have shown that it is possible
to take an existing interpreter in C and replace relevant parts of it with
RPython, creating a meta-tracing VM. To the best of our knowledge, the first
time this has been done. It may be possible to apply this technique to other
systems (including non-DBMSs), with minor adjustments.

\textbf{Acknowledgements:} This research was funded by the EPSRC Cooler
(EP/K01790X/1) grant and Lecture (EP/L02344X/1) fellowship. We thank
Geoff French for comments.

\bibliographystyle{plainurl}


\bibliography{bolz}

\appendix

\section{Unported Aspects of SQLite}
\label{sec:stuff we didn't port}

When porting SQLite C code to RPython, we did not port the following aspects:

\begin{itemize}
\item Assert statements, which are removed by the C compiler.
\item Statements related to tests and debugging, which expand to nothing in
    production builds:
\begin{itemize}
    \item \texttt{VdbeBranchTaken}
    \item \texttt{REGISTER\_TRACE}
    \item \texttt{SQLITE\_DEBUG}
    \item \texttt{memAboutToChange}
    \item \texttt{UPDATE\_MAX\_BLOBSIZE}
\end{itemize}
\item Blocks of \lstinline{#ifdef} and \lstinline{#ifndef}, which are usually
not included in default production builds:
\begin{itemize}
    \item \texttt{SQLITE\_DEBUG}
    \item \texttt{SQLITE\_OMIT\_FLOATING\_POINT}
\end{itemize}
\item We assumed \texttt{SQLITE\_THREADSAFE} to be false, which SQLite
    recommends for best single-threaded performance.
\item We decided to compile and port with
    \texttt{SQLITE\_OMIT\_PROGRESS\_CALLBACK} turned on. Usually SQLite makes it
    possible to register a progress callback that is called every $n$ opcodes.
    We plan to implement this in the future. Note that in our evaluation,
    we compared SQPyte to SQLite with callbacks similarly omitted, thus 
    ensuring an apples-to-apples comparison.
\end{itemize}

\end{document}